\documentclass[prb,floatfix,twocolumn,letterpaper,nobalancelastpage]{revtex4-2}

\usepackage{mycommands}

\begin{document}
    
    \title{Fractional Chern insulators with a non-Landau level continuum limit}
    
    \author{David Bauer}
    \affiliation{Department of Physics and Astronomy, University of California at Los Angeles, 475 Portola Plaza, Los Angeles, California 90095, USA}
    
    \author{Spenser Talkington}
    \affiliation{Department of Physics and Astronomy, University of California at Los Angeles, 475 Portola Plaza, Los Angeles, California 90095, USA}
    
    \author{Fenner Harper}
    \affiliation{Department of Physics and Astronomy, University of California at Los Angeles, 475 Portola Plaza, Los Angeles, California 90095, USA}
    
    \author{Bartholomew Andrews}
    \affiliation{Department of Physics and Astronomy, University of California at Los Angeles, 475 Portola Plaza, Los Angeles, California 90095, USA}
    
    \author{Rahul Roy}
    \email{rroy@physics.ucla.edu}
    \affiliation{Department of Physics and Astronomy, University of California at Los Angeles, 475 Portola Plaza, Los Angeles, California 90095, USA}
    
    \date{\today}
    \begin{abstract}
        Recent developments in fractional quantum Hall (FQH) physics highlight the importance of studying FQH phases of particles partially occupying energy bands that are not Landau levels. FQH phases in the regime of strong lattice effects, called fractional Chern insulators, provide one setting for such studies. As the strength of lattice effects vanishes, the bands of generic lattice models asymptotically approach Landau levels. In this article, we construct lattice models for single-particle bands that are distinct from Landau levels even in this continuum limit. We describe how the distinction between such bands and Landau levels is quantified by band geometry over the magnetic Brillouin zone and reflected in the electromagnetic response. We analyze the localization-delocalization transition in one such model and compute a localization length exponent of 2.57(3). Moreover, we study interactions projected to these bands and find signatures of bosonic and fermionic Laughlin states. Most pertinently, our models allow us to isolate conditions for optimal band geometry and gain further insight into the stability of FQH phases on lattices.
    \end{abstract}
    
    \maketitle
    
    \section{Introduction}
    \label{sec-intro}
    
        The incompressible liquid phases of the fractional quantum Hall effect (FQHE) serve as prototypical examples of topologically ordered or gapped quantum liquid phases in 2+1 dimensions~\cite{yoshioka_quantum_2002,fradkin_field_2013}. These phases are characterized by their long-range entanglement structure, which gives rise to topological quasiparticles and universal, quantized linear responses. Theoretical models of the quantum Hall (QH) fluid---based, for example, on model wavefunctions~\cite{laughlin_anomalous_1983}---often make the simplifying assumption that the microscopic constituent particles occupy states in a single Landau level. Some properties of Landau levels, such as nonzero Chern number \cite{thouless_quantized_1982}, are essential for the quantum Hall effect (QHE), but the existence of fractional Chern insulators (FCIs)---time-reversal (TR) symmetry-breaking, fractionalized phases observed in the regime of strong lattice effects~\cite{Sheng2011, Bergholtz:2013ue,parameswaran_fractional_2013,neupert2015fractional}---indicates that Landau levels are in general not required.
        
        Recently, FCIs have been experimentally observed via compressibility measurements in a set of two-dimensional moir{\'e} systems~\cite{Spantoneaan8458,xie_fractional_2021}. Moreover, efforts to realize FCIs at zero magnetic field are still ongoing and rely on engineering the optimal band conditions for fractionalized states on lattices. FCI phases show a wide range of stability dependent on several factors, including the band flatness and strength of interactions, which were identified early on~\cite{Bergholtz:2013ue}, as well as band geometry, which corresponds to quantities such as the Berry curvature and Fubini-Study (FS) metric~\cite{parameswaran_fractional_2013}. Under a set of conditions that can be imposed on the band geometry, a Chern band can resemble the lowest Landau level, which is believed to yield ``ideally" stable FCI phases~\cite{Lee2017,wang2021exact,wang2021hierarchy,varjas2021topological,mera2021engineering}. The relevance of these conditions has been investigated in numerical simulations~\cite{jackson_geometric_2015,Claassen2015,bauer_quantum_2016,andrews2018stability,andrews2021stability}, but not so far in models where the role of the conditions can be isolated from one another.
        
        Motivated by these experimental and theoretical challenges, in this paper, we introduce a set of models with no quadratic term in their momentum-space Hamiltonian, and show that they have a continuum limit that differs from Landau levels. Since these models allow us to tune the FS metric while keeping the Berry curvature constant, we also probe the role of subsets of the band geometric conditions in the stability of fractionalized topological phases. We focus on one family of these models and study single-particle properties, including the details of the band geometry, the electromagnetic response, and the localization-delocalization transition in the presence of disorder. Moreover, we demonstrate and analyze the stability of FQH phases in these bands and discuss the scope of the ``geometric stability hypothesis" (GSH)~\cite{jackson_geometric_2015}. These results introduce an alternative realization of the FQHE, which resides in continuum non-Landau levels, and contribute to the current research on engineering correlated states using band geometry~\cite{wang2021hierarchy, xie_fractional_2021, varjas2021topological, mera2021engineering, rossi2021quantum}.
        
        The body of this article is organized in five sections. We begin in Sec.~\ref{sec-prelim} with the observation that Harper-Hofstadter tight-binding models in the continuum limit of small flux per plaquette generically lead to a perturbative, mean-field Hamiltonian that reproduces the Landau level Hamiltonian. In Sec.~\ref{section-single-particle}, we then exhibit a series of such tight-binding models that do not have Landau levels as their continuum eigenstates and quantify the distinction between these bands and their Landau level counterparts using Chern band geometry. In Sec.~\ref{sec-em-response}, we apply a recently-developed technique, based on elementary perturbation theory~\cite{harper_finite-wavevector_2018}, for calculating current responses of Chern bands to spatially-inhomogeneous electric fields. This calculation yields non-universal corrections to the finite-wavevector conductivity, in contrast to isotropic continuum models, which have a universal term related to the Hall viscosity. In Sec.~\ref{sec-localization}, we consider the localization-delocalization transition in the lowest level of the model with quenched onsite disorder. Using a transfer matrix method, we find that the localization-length critical exponent for the disordered model is $2.57(3)$.  We discuss the relationship of this exponent with existing studies of localization-delocalization transitions. Finally, in Sec.~\ref{sec-geometry}, we use numerical exact diagonalization to study the effect of interactions in these bands and find numerical signatures of Laughlin states of bosons and fermions. We study the relationship between the geometry of these non-Landau bands and the stability of FQH phases of particles occupying such  bands, further extending the GSH for Chern bands~\cite{jackson_geometric_2015}.
        
    \section{Preliminary Discussion}
    \label{sec-prelim}
            
        In this section, we review how the Landau level limit can be recovered from the Harper-Hofstadter model in Sec.~\ref{landau-level-limit} and discuss the quantum geometry of Chern bands in Sec.~\ref{chern-band-geometry}.    
        
        \subsection{Effective Landau levels from weak-field Harper-Hofstadter models}
        \label{landau-level-limit}
        
            We start by considering the weak-field limit $\phi \rightarrow 0$ of a Harper-Hofstadter tight-binding Hamiltonian, and show how, to lowest order in $\phi$, this Hamiltonian reproduces the Landau level Hamiltonian.
            
            We introduce a uniform background magnetic field $B$ perpendicular to the spatial extent of the lattice and choose the value of $B$ such that the flux per lattice plaquette is $\phi = Ba^2 = \frac{p}{q}\phi_0$, where $\phi_0 = 2\pi \hbar /e$ is the magnetic flux quantum and $p$ and $q$ are coprime integers. We will mostly consider the case $p=1$, in which the band structure is conceptually simplest~\cite{Harper:2014vi,Schoonderwoerd2019}. In terms of the magnetic length $\ell = \sqrt{\hbar/eB}$, and lattice spacing $a$, $\phi = \hbar a^2/(e \ell^2)$. For the rest of this article, we will work in units where $\hbar/e= 1$, so that $\phi = a^2/\ell^2$ is a dimensionless ratio of area scales.
            
            In the presence of the magnetic field, the na\"ive translation operators do not transform correctly under gauge transformations~\cite{fradkin_field_2013}, and we must accompany translations by compensatory gauge transformations~\cite{zak_magnetic_1964}. The translation operators with the appropriate transformation properties are a sum over lattice sites $\bm{m}$,
            \begin{align}
            T_a = \sum_{\bm{m}} e^{i\theta_a(\bm{m})} c^{\dag}_{\bm{m} + \bm{e}_a}c_{\bm{m}},
            \end{align}
            where the phases $e^{i\theta_a(\bm{m})}$ satisfy 
            \begin{align}
            \theta_1(\bm{m}) + \theta_2(\bm{m} + \bm{e}_1) - \theta_1(\bm{m} + \bm{e}_2) - \theta_2(\bm{m}) = \phi.
            \end{align}
            The components of $\bm{T}$ do not commute, but satisfy
            \begin{align}
            T_x T_y = \exp(i\phi) T_y T_x.
            \end{align}
            The lattice translation operators $T_a$ are unitary, so we can write them in terms of Hermitian generators $T_a = \exp(i K_a)$. The $K_a$ are the lattice analogues of the covariant momentum operators $\pi_a$, and we will sometimes call them momenta for brevity. These operators satisfy the commutator
            \begin{align}
            \comm{K_x}{K_y} = \phi.
            \end{align}
            
            The non-commutativity of the $T_a$ leads to an ambiguity in constructing the most general tight-binding Hamiltonian when $B\neq0$. We resolve this ambiguity by specifying that the operator for hopping $j$ sites in the $x$-direction and $k$ sites in the $y$-direction be the symmetric sum $T_x^j T_y^k + T_y^k T_x^j$, so that our tight-binding Hamiltonian is
            \begin{align}
            H_{\text{TB}} = -\sum_{j,k} t_{jk}\left(T_x^j T_y^k + T_y^k T_x^j\right) + \text{H.c.}.
            \end{align}
            We could also have resolved this ambiguity by a gauge choice, although for now we maintain gauge symmetry. We note that in our notation, the hopping amplitudes $t_{jk}$ are always real, and any complex phase factors come from the action of the translation operators.
            
            Let us look at the nearest-neighbor (NN) hopping Hamiltonian containing only first powers of the translation operators,
            \begin{align}
            H_{\text{NN}} = -t_{10}\left(T_x + T_x^{\dag}\right) - t_{01}\left(T_y + T_y^{\dag}\right).
            \end{align}
            In the $C_4$-symmetric case, $t_{01} = t_{10}$, $H_{\text{NN}}$ is the Hamiltonian of the familiar Hofstadter model with non-zero amplitude only for NN hoppings. Since $T_a = \exp(iK_a)$,
            \begin{align}
            H_{\text{NN}}= -2t_{10}\cos\left(K_x\right) - 2 t_{01}\cos\left(K_y\right).
            \end{align}
            In order to make the dependence on $\phi$ explicit, we rescale the $K_a$ operators, defining $\sqrt{\phi} P_a = K_a$. Expressed in terms of the $P_a$ operators the tight-binding model is
            \begin{align}
            \begin{split}
            H_{\text{NN}} &= -2\sum_{n=0}^{\infty}  \frac{(-1)^{n}\phi^{n}}{(2n)!} \left(t_{10} P^{2n}_x + t_{01}P^{2n}_y\right)\\
            &= -2 + 2\phi\frac{\left(t_{10} P^{2}_x + t_{01} P^{2}_y\right)}{2} +O(\phi^2).
            \end{split}
            \end{align}
            Now let $t_{01} = \alpha^2 t_{10} = \alpha^2 t$, i.e., $t$ is a common hopping energy scale and $\alpha$ parameterizes anisotropy in the hopping amplitudes. Then to lowest order in $\phi$, we have an effective Hamiltonian $H_{\text{eff}}= t\phi \left(P^{2}_x + \alpha^2P^{2}_y\right)$. We can rewrite this in terms of momentum operators that satisfy $\comm{\pi_x}{\pi_y} = i\hbar e B$ as
            \begin{align}
            H_{\text{eff}} = \frac{1}{2m^*}\left(\pi_x^2 + \pi_y^2\right),
            \end{align}
            showing that our effective Hamiltonian is isomorphic to the Landau level Hamiltonian with effective mass $m^* = \hbar^2/(2ta^2\alpha)$. In order to recover all of the physics of Landau levels, we also need an analogue of the guiding-center operators that commute with the Hamiltonian but not with one another, leading to an extensive degeneracy. Here, this role is filled by the magnetic translation operators, which we define in the following section. If we consider not just NN but also longer range hopping terms, the details of the above argument are slightly more complicated, but to lowest order the Hamiltonian remains quadratic in the momenta. This gives us an effective mean-field Hamiltonian for the cyclotron degrees of freedom. 

        \subsection{Chern band geometry}
        \label{chern-band-geometry}
        
            Instead of the continuous translation symmetry that generates degenerate states in Landau levels in infinite plane and toroidal geometries, our system has discrete translation symmetry corresponding to the magnetic translation operators $U_a$. If we define a magnetic unit cell (MUC) with dimensions $X_{\text{MUC}} \times Y_{\text{MUC}} = A_{\text{MUC}}$ enclosing flux $\phi=1$, translations by one MUC in either direction commute with each other and the Hamiltonian. The simultaneous eigenstates of the Hamiltonian and MUC translations are $\ket{n,\bm{k}}$, where $\bm{k}$ takes values in the magnetic Brillouin zone and $n$ labels eigenstates of $H$. We can write the Hamiltonian
            \begin{align}
            \label{band-projector-ham}
            H = \sum_{n}\int d^2k\, E_{n}(\bm{k}) \mathcal{P}_{n,\bm{k}},
            \end{align}
            where $E_n$ is the dispersion of the $n$-th band of the Hamiltonian, and $\mathcal{P}_{n,\bm{k}}$ is the projector onto the state $\ket{n,\bm{k}}$.
            
            Unlike in Landau levels, the dispersion $E_{n}(\bm{k})$ generically has some finite width, and the states in the band are not exactly degenerate. In the weak-field case this bandwidth will be exponentially small~\cite{Harper:2014vi}, while in the strong-field limit we may follow the usual procedure of adding long-range hoppings to flatten the band~\cite{Bergholtz:2013ue,parameswaran_fractional_2013}. In either case, we will neglect the bandwidth. Despite being dispersionless, these Chern bands are distinct from Landau levels. We can quantify this distinction by considering the Berry curvature and FS metric defined on the magnetic Brillouin zone~\cite{parameswaran_fractional_2013,roy_band_2014,Claassen2015}. (The FS metric has also been called the Bures metric in this context \cite{palumbo_momentum-space_2017}.) For notational simplicity, we define $\partial_a = \partial_{k_a}$. Then, respectively, the Berry curvature and FS metric components for a band with projector $\mathcal{P}_{\bm{k}}$ are
            \begin{align}
            \B(\bm{k}) = \epsilon_{ab}\text{Tr}\left(\partial_{a}\mathcal{P}_{\bm{k}}\partial_b \mathcal{P}_{\bm{k}}\right),
            \end{align}
            and, defining the orthogonal projector $\mathcal{Q}_{\bm{k}} = \bm{1} - \mathcal{P}_{\bm{k}}$,
            \begin{align}
            g_{ab}(\bm{k}) = \frac{1}{2}\text{Tr}\left(\partial_{a}\mathcal{Q}_{\bm{k}}\partial_{b}\mathcal{P}_{\bm{k}} + \partial_{b}\mathcal{Q}_{\bm{k}}\partial_{a} \mathcal{P}_{\bm{k}}\right).
            \end{align}
            The traces in these expressions are taken over the cyclotron Hilbert space. The TR-symmetry breaking Chern bands that we want to study are those with nonvanishing first Chern number
            \begin{align}
            C_1 = \int d^2k\, \B(\bm{k}).
            \end{align}
            
            The FS metric and Berry curvature for any band obey the inequalities~\cite{roy_band_2014}
            \begin{align}
            \begin{split}
            \label{berry-metric-inequalities}
            \text{Det}\,g(\bm{k})&\geq\frac{1}{4}|\B(\bm{k})|^2,\\
            \text{Tr}\,g(\bm{k})&\geq|\B(\bm{k})|,
            \end{split}
            \end{align}
            which we call the determinant inequality and trace inequality, respectively. For Landau levels, these inequalities are saturated. The degree to which the FS metric and Berry curvature for a particular band saturate these inequalities therefore provides a quantitative measure of deviations from Landau level behavior and may also provide a measure of stability of FQH phases in the band. We define
            \begin{align}
            \begin{split}
            \label{geometry-inequalities}
            \mathcal{D}(\bm{k}) &= \text{Det}\,g(\bm{k})-\frac{1}{4}|\B(\bm{k})|^2,\\
            \mathcal{T}(\bm{k}) &=\text{Tr}\,g(\bm{k})-|\B(\bm{k})|,
            \end{split}
            \end{align}
            measuring the degree of saturation, and we refer to these quantities as the determinant inequality saturation measure (DISM) and trace inequality saturation measure (TISM), respectively.
            
            \begin{figure}[bh]
            \centering
            \includegraphics[width=3.1in]{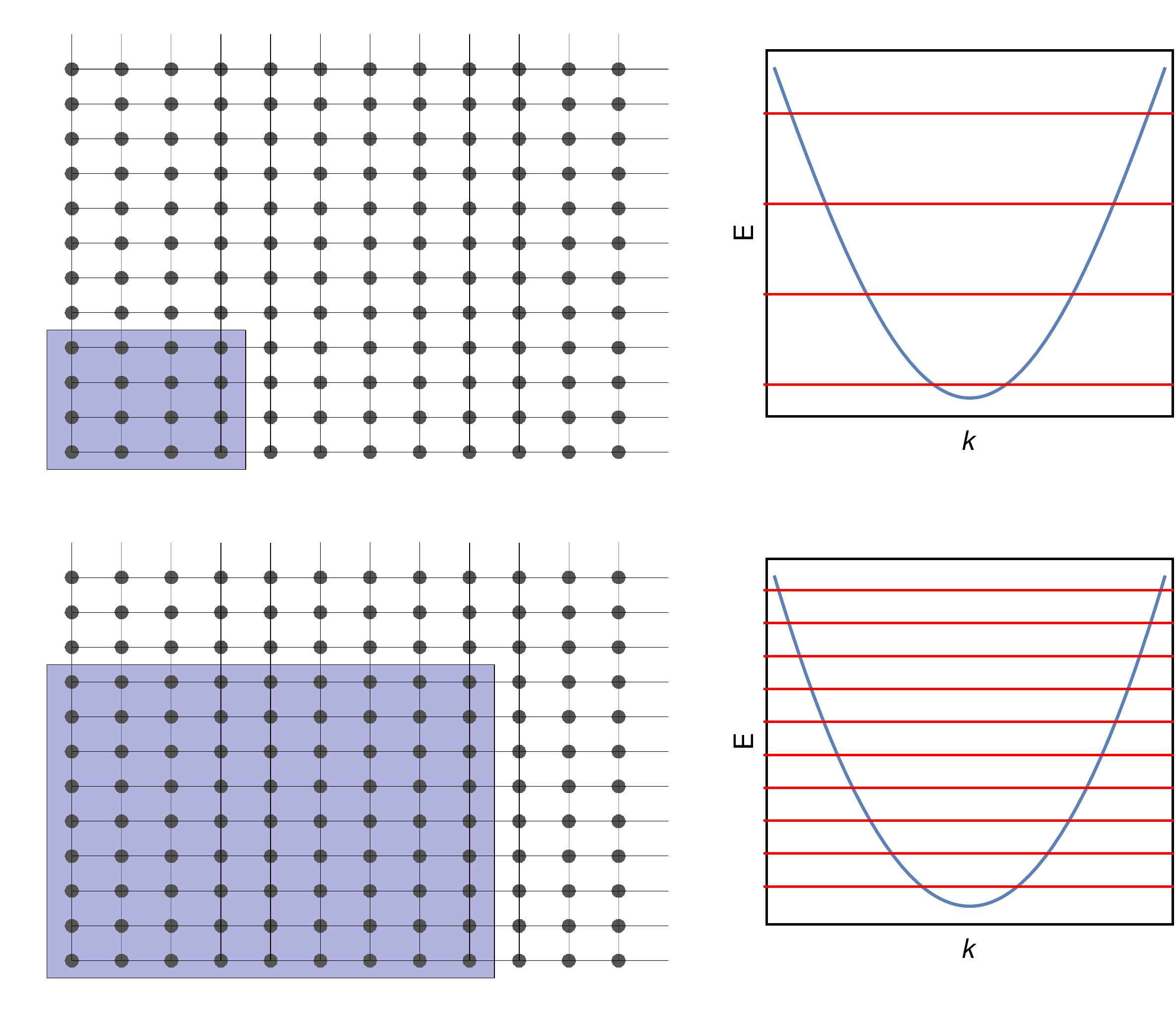}
            \caption{\label{bands-schematic}Schematic depiction of Landau levels as the weak-field limit of Harper-Hofstadter bands near the minimum of a periodic potential. As the flux per plaquette is decreased, the elementary magnetic unit cell---represented by the blue region of the lattice---must increase in size to enclose the same number of flux quanta. In turn, the number of states in each band decreases, and each band comprises states increasingly close to the minimum of the zero-field dispersion.}
            \end{figure}
            
            The weak-field/continuum limit is implemented in the $\bm{k}$-representation by expanding the band dispersion $E_{n}(\bm{k})$ near a band minimum. The lowest-order term in such an expansion will generically be quadratic in $\bm{k}$, which essentially follows from our argument in the previous section. Since the mean electron density $\rho_0$ in QH systems is pegged to the magnetic flux density by the filling fraction
            \begin{align}
            \nu = \frac{N_p}{N_s} = N_p\frac{A_{\text{tot}}}{A_{\text{MUC}}},
            \end{align}
            the weak-field limit corresponds to the limit of a dilute electron fluid. This leads to a more conceptual picture, shown in cartoon form in Fig.~\ref{bands-schematic} for how decreasing flux per plaquette leads to continuum limit Landau levels. If we fix the overall system size and the magnetic flux attached to each guiding-center lattice site (the flux through each MUC), decreasing flux per plaquette must increase the area of each MUC. We therefore increase the number of states in the cyclotron Hilbert space, while decreasing the number of guiding-center states available within each band. The states that live near the band minimum when there is no magnetic field then redistribute into increasingly higher bands as $\phi$ decreases.

    \section{Non-Landau effective Hamiltonians}
    \label{section-single-particle}
    
        In this section, we introduce and study a new class of weak-field effective Hamiltonians that cannot be transformed into the Landau Hamiltonian at lowest order in $\phi$. We derive a family of such models and study their band geometry in Sec.~\ref{subsec:zero_quadratic}, we analyze the non-perturbative corrections due to the lattice in Sec.~\ref{subsec:non_pert}, we investigate the effect of explicitly removing the Landau level Hamiltonian from our models in Sec.~\ref{subsec:no_h0}, and we consider the effect of anisotropy in Sec.~\ref{subsec:aniso}.
        
        \subsection{Zero-quadratic models}
        \label{subsec:zero_quadratic}
        
        To start, we consider the following tight-binding Hamiltonian obtained by adding a next-nearest-neighbor (NNN) hopping to the Hofstadter model:
        \begin{align}
        \label{quartic-harper}
        H_{\text{NNN}} = &-t_1 \left(T_x + T_x^{\dag} + T_y + T_y^{\dag}\right)\nonumber\\ &- t_2 \left(T_x^{2} + T_x^{\dag 2} + T_y^{2} + T_y^{\dag 2}\right).
        \end{align}
        For this particular model, we have omitted the NNN hopping diagonally across the elementary plaquette. For a derivation with all NNN hoppings that respect the $C_4$ symmetry, we refer the reader to Appendix~\ref{allNNN}. Unless explicitly stated, we will assume from here onward that $t_1$ sets the overall scale of the Hamiltonian, and set $t_1=1$; the remaining hopping amplitudes in each expression should then be understood as dimensionless ratios.
        
        As in Sec.~\ref{landau-level-limit}, we write this in terms of the Hermitian generators of lattice translations,
        \begin{align}
        H_{\text{NNN}} = &-2\left(\cos(K_x) + \cos(K_y)\right)\nonumber\\ &- 2t_2\left(\cos(2K_x) + \cos(2K_y)\right).
        \end{align}
        Replacing the cosine terms by their Taylor expansion, the terms lowest order in the momenta are 
        \begin{align}
        H_{\text{NNN}} = &-4 - 4 t_2 + (1 + 4t_2) \left(K_x^2 + K_y^2\right)\nonumber \\
        &- \left(\frac{1}{12} + \frac{4}{3}t_2\right) \left(K_x^4 + K_y^4\right) + \ldots.
        \end{align}
        If we make the particular choice of hopping amplitudes $t_2 = -1/4$, then the quadratic terms vanish exactly, and we are left with an effective Hamiltonian that is quartic in the momenta to lowest order,
        \begin{align}
        \label{Hamiltonian-quartic-effective}
        H_{\text{eff}} = -3 + \tfrac{1}{4} \left(K_x^4 + K_y^4\right).
        \end{align}
        We will refer to this tight-binding model, Eq.~\eqref{quartic-harper} with appropriate $(t_1,t_2)$, as the \emph{zero-quadratic model}. The negative hopping amplitude necessary to eliminate the quadratic term may be realized in optical lattice experiments by periodic shaking of the lattice~\cite{eckardt_colloquium_2017}. Unlike the Landau level Hamiltonian, this Hamiltonian does not have $SO(2)$ rotational symmetry, but it is symmetric under the square lattice point group $D_4$. We note that the particular quartic momentum operator in \eqref{Hamiltonian-quartic-effective} can be written as
        \begin{align}
        K_x^4 + K_y^4 = \left(K_x^2 + K_y^2\right)^2 - \left(K_x^2K_y^2 + K_y^2K_x^2\right),
        \end{align}
        that is, as the square of the Landau level Hamiltonian plus a term that explicitly breaks continuous rotational symmetry.
        
        \begin{figure}[thb]
        \centering
        \hspace{-0.25in}\includegraphics[width=3.0in]{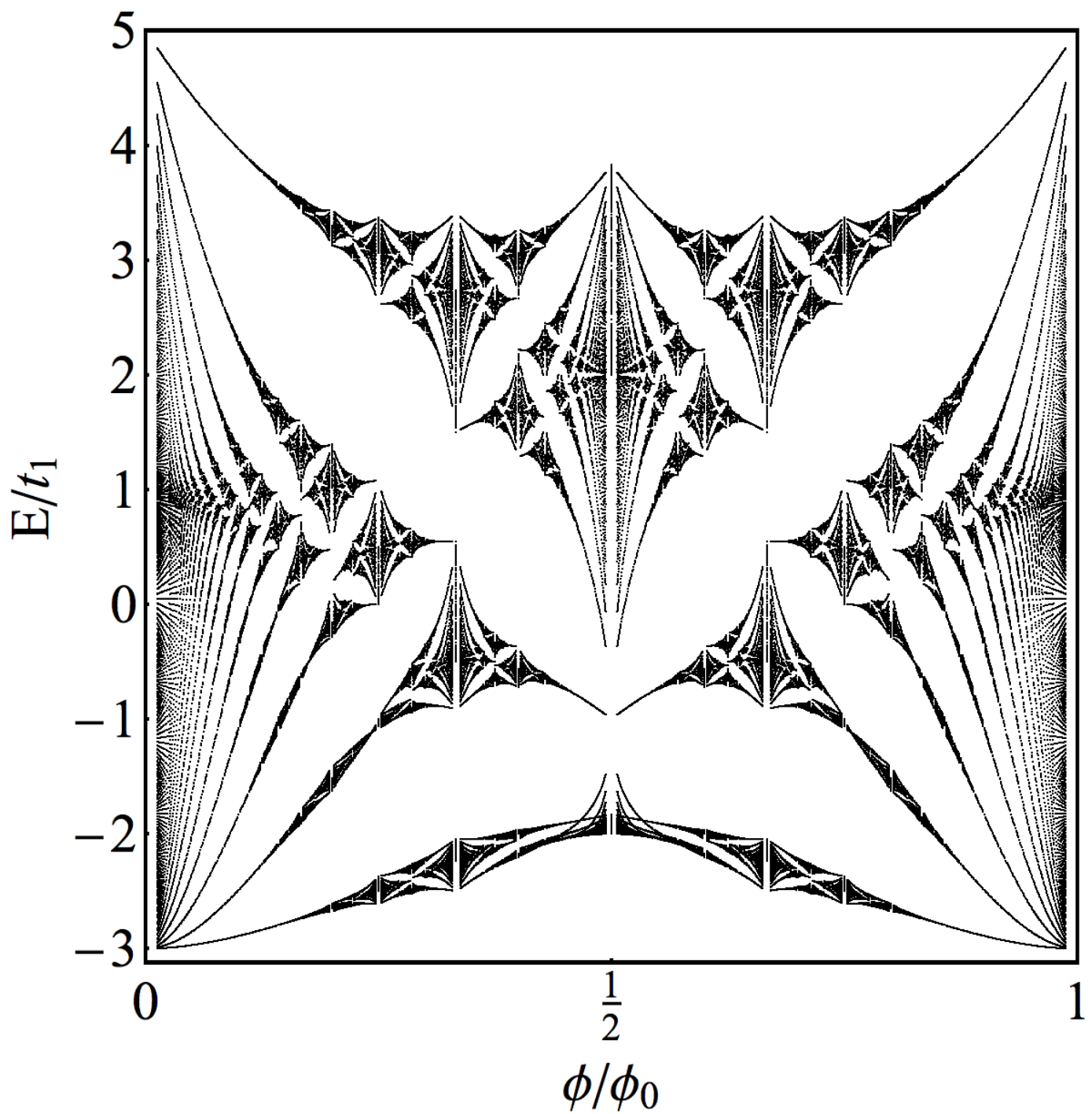}
        \caption{\label{butterfly-plot} Energy eigenvalues of the zero-quadratic model, Eq.~\eqref{quartic-harper} with $t_1 = 1$, $t_2 = -1/4$, as a function of magnetic flux per elementary lattice plaquette $\phi = (p/q)\phi_0$.}
        \end{figure}
        
        We now consider some details of the model given by Eq.~\eqref{quartic-harper}, in particular its continuum limit. We plot the energy eigenvalues obtained from the Harper-type equation for this model as a function of $\phi$---the analogue of the Hofstadter ``butterfly'' for this model~\cite{hofstadter_energy_1976}---and display this in Fig.~\ref{butterfly-plot}. In the Hofstadter model, the bands are approximately linear in $\phi$ at low flux. In the zero-quadratic model, the bands are approximately quadratic in $\phi$ at low flux, as we quantify later. We compute the Berry curvature and Chern number of these bands for $\phi = \frac{1}{N}$ and verify that they have Chern number $|C_1| = 1$.
        
        As for Landau levels, we can study the spectrum of the effective Hamiltonian by introducing ladder operators $a$ and $a^{\dag}$ corresponding to the cyclotron momenta $K_a$. Here, we choose the operators
        \begin{align}
        a &= \frac{1}{\sqrt{2\phi}}\left(K_x - iK_y\right),\\
        a^{\dag} &= \frac{1}{\sqrt{2\phi}}\left(K_x + iK_y\right),
        \end{align}
        with inverse map
        \begin{align}
        K_x &= \sqrt{\frac{\phi}{2}}\left(a + a^{\dag}\right),\\
        K_y &= -i\sqrt{\frac{\phi}{2}}\left(a - a^{\dag}\right).
        \end{align}
        We also define the dimensionless operator
        \begin{align}
        h_0 = \left(a^{\dag}a + \frac{1}{2}\right).
        \end{align}
        In terms of these operators,
        \begin{align}
        \left(K_x^2 + K_y^2\right)^2 &= 4\phi^2 h_0^2, \\
        \left(K_x^2K_y^2 + K_y^2K_x^2\right) &= \phi^2\left(h_0^2 -\frac{1}{2}\left(a^4 + a^{\dag\,4}\right) - \frac{3}{4}\right),
        \end{align}
        and the effective Hamiltonian is
        \begin{align}
        \label{effective-fock-Hamiltonian}
        H_{\text{eff}} = \frac{\phi^2}{8}\left[\left(a^4 + a^{\dag 4}\right) + 6h_0^2 + \frac{3}{2}\right].
        \end{align}
        
        We numerically approximate this Hamiltonian by working in a basis of number eigenstates $\ket{n}$ satisfying $a^{\dag}a{\ket{n}}=n{\ket{n}}$ and truncating to a finite-dimensional subspace. This gives estimates for the cyclotron energies and overlaps with the Landau level states. We find good agreement between this continuum approximation truncated to $n \leq 1000$ and exact numerical energy levels of the lattice Hamiltonian for small $\varepsilon$. The first two nonzero overlaps of the ground state $\ket{\tilde{0}}$ of the Hamiltonian \eqref{effective-fock-Hamiltonian} with the Landau level states are $\braket{\tilde{0}}{0}\approx0.9991$, $\braket{\tilde{0}}{4}\approx-0.0422$. When expressed with ladder operators, the TISM takes the particularly simple form $\expval{\mathcal{T}} = 2\expval{a^{\dag}a}$~\cite{bauer_quantum_2016}; that is, $\expval{\mathcal{T}}$ is twice the mean LL occupation number $n_0$. Calculating this in the truncated Landau level basis, we find $\expval{\mathcal{T}} \approx0.0143$, in good agreement with the value found from integrating the lattice $\mathcal{T}(\bm{k})$ over the MBZ, $\expval{\mathcal{T}} \approx 0.0145$.
        
        We also study the spectrum of cyclotron orbits of this Hamiltonian semiclassically by applying the Bohr-Sommerfeld quantization condition that the adiabatic invariants of the classical Hamiltonian be quantized. In our notation, this condition takes the form
        \begin{align}
        \oint\limits_{H=E_n} K_x\, dK_y = 2\pi n,
        \end{align}
        with the integral taken over a closed curve of constant energy in classical phase space. From this condition we find
        \begin{align}
        E_n \sim n^2\phi^2, 
        \end{align}
        in agreement both with the numerically-obtained, approximate spacing of the cyclotron levels, and with the quadratic dependence of $E$ on $\phi$ observed in the butterfly plot, Fig.~\ref{butterfly-plot}.
        
        \begin{figure}[thb]
        \centering
        \includegraphics[width=3.0in]{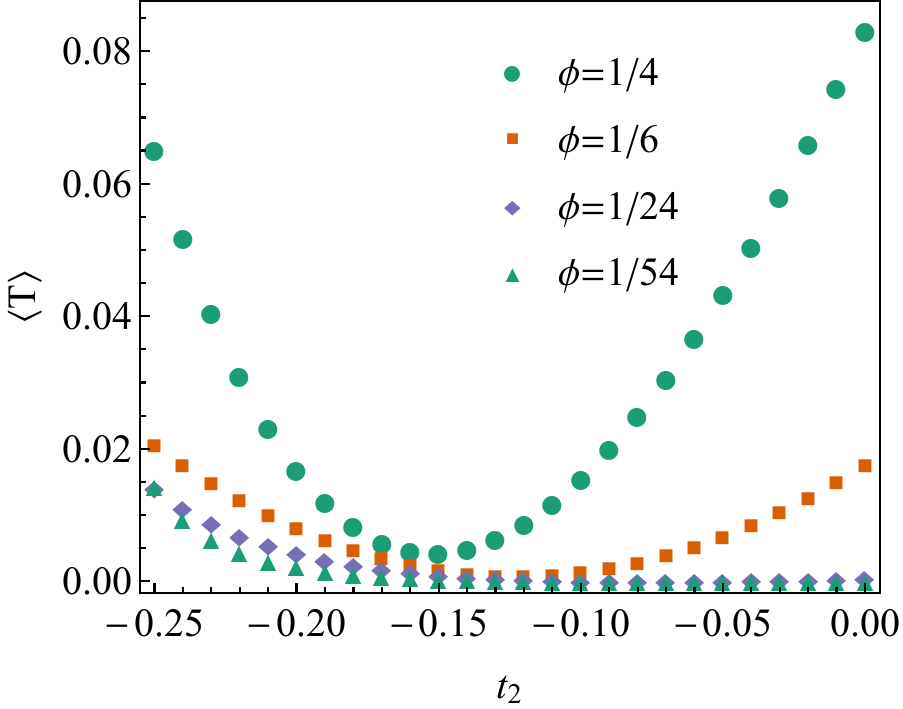}
        \caption{\label{trace-delta-plot}TISM $\expval{\mathcal{T}}$ for the lowest-lying band of \eqref{quartic-harper} as a function of $t_2$ with $t_1 = 1$. For $t_2 = 0$ this is the Hofstadter model, while $t_2 = -1/4$ gives the zero-quadratic model with effective Hamiltonian~\eqref{Hamiltonian-quartic-effective}.}
        \end{figure}
        
        We now consider general values of $t_2$. In this case, the effective Hamiltonian is, up to a constant,
        \begin{align}
        \label{t2-Hamiltonian}
        H_{\text{eff}} = &-\frac{1 + 16t_2}{12}\left(K_x^4 + K_y^4\right)\nonumber\\ &+ \left(1 + 4t_2\right) \left(K_x^2 + K_y^2\right),
        \end{align}
        or
        \begin{align}
        H_{\text{eff}} = 2&\phi\left(1 + 4t_2\right) h_0\nonumber\\  - &\phi^2\frac{\left(1 + 16t_2\right)}{4} \left(h_0^2 + \frac{\left(a^{4} + a^{\dag 4}\right)}{6} + \frac{1}{4}\right).
        \end{align}
        For any value of $t_2$ away from the fine-tuned point $t_2 = -1/4$, we may always choose $\phi$ small enough that the quadratic term dominates and we recover the Landau level Hamiltonian. However, if we consider $\phi$ small but fixed, we can make the weight of the quadratic term small by tuning $t_2$ appropriately. In this regime, we can treat the quadratic term in the Hamiltonian as a perturbation to the quartic term, with perturbative parameter
        \begin{align}
        \varepsilon = \frac{8}{\phi}\frac{\left(1 + 4t_2\right)}{\left(1 + 16t_2\right)}.
        \end{align}
        A weak upper bound on the regime in which we may treat the quadratic term as a perturbation is given by setting $\varepsilon < 1$ or
        \begin{align}
        -\frac{1}{16} > t_2 > -\frac{1}{4} - \frac{3}{32}\phi +O(\phi^2).
        \end{align}
        In Fig.~\ref{trace-delta-plot}, we plot $\expval{\mathcal{T}}$ for the lowest band of this Hamiltonian for various values of $t_2$, including values that interpolate between the Hofstadter model and our zero-quadratic model. For large values of $\phi$, this plot shows a clear local minimum, which we interpret to be the point where the weight of $h_0$ and its powers are maximized. As $\phi$ decreases, $\expval{\mathcal{T}}$ flattens toward zero for $t_2$ near the Hofstadter regime, in line with asymptotic vanishing of $\expval{\mathcal{T}}$ in the Hofstadter model~\cite{bauer_quantum_2016}. The finite value of $\expval{\mathcal{T}}$ for the zero-quadratic model in the limit of small flux density highlights the convergence to non-Landau levels.
        
        \begin{figure}[thb]
        \centering
        \includegraphics[width=3.0in]{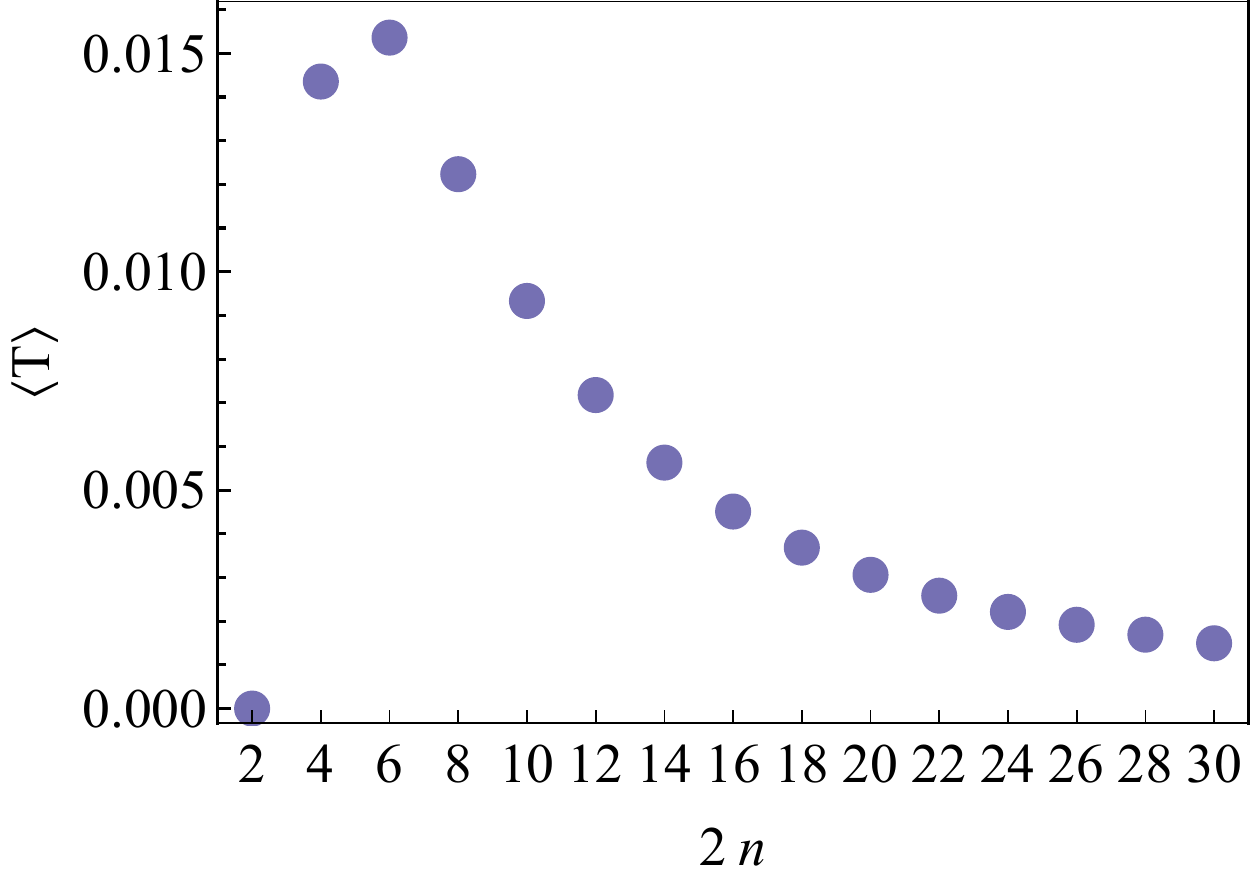}
        \caption{\label{plot-2n-trace}TISM $\expval{\mathcal{T}} = 2 \expval{a^{\dag} a}$ for the ground state band of the Hamiltonian $H^{(2n)}=K_x^{2n} + K_y^{2n}$. When $n=1$, this is the Landau level Hamiltonian for which the lowest-lying band has $\expval{\mathcal{T}}=0$ as shown.}
        \end{figure}
        
        By introducing additional hopping amplitudes between non-neighboring sites and tuning the amplitudes appropriately, we may eliminate increasingly higher-order terms in the Hamiltonian and generate other zero-quadratic models with different effective Hamiltonians. We focus on the case in which we add only straight-line hoppings, as we did for the zero-quadratic model in Eq.~\eqref{quartic-harper}. For example, we obtain the hexic effective Hamiltonian
        \begin{align}
        H_{\text{eff}} = -\frac{4}{3} + \frac{K_x^6 + K_y^6}{15}
        \end{align}
        by choosing $t_{20}=t_{02}=-2/5$ and $t_{30}=t_{03}=1/15$, and so on, generalizing to the case
        \begin{align}
        H^{(2n)}_{\text{eff}} = K_x^{2n} + K_y^{2n}.
        \end{align}
        We find the perturbative, finite-size eigenstates of these effective Hamiltonians numerically, and display the TISM of their ground states in Fig.~\ref{plot-2n-trace}. We find that the TISM is maximum for $2n=6$, which shows the largest deviation from Landau levels in the continuum limit, and that for larger values of $n$ it decreases monotonically at least until $2n=30$.
        
        \subsection{Non-perturbative corrections and uniformity of band geometry}
        \label{subsec:non_pert}
        
            In the perturbative treatment we have employed so far, we expand about a band minimum to obtain an effective mean-field Hamiltonian. While this method gives corrections to the energy levels due to changes in shape of the effective dispersion near the minimum, it neglects corrections arising from the discreteness of the lattice. In particular, since Landau levels have a perfectly uniform energy dispersion, Berry curvature, and FS metric, any fluctuations in these must come from non-perturbative corrections. These non-perturbative corrections arise from tunneling between band minima, so intuitively we may expect them to decay exponentially in the width of the potential barrier. 
            
            \begin{figure}[thb]
            \centering
            \includegraphics[width=3.1in]{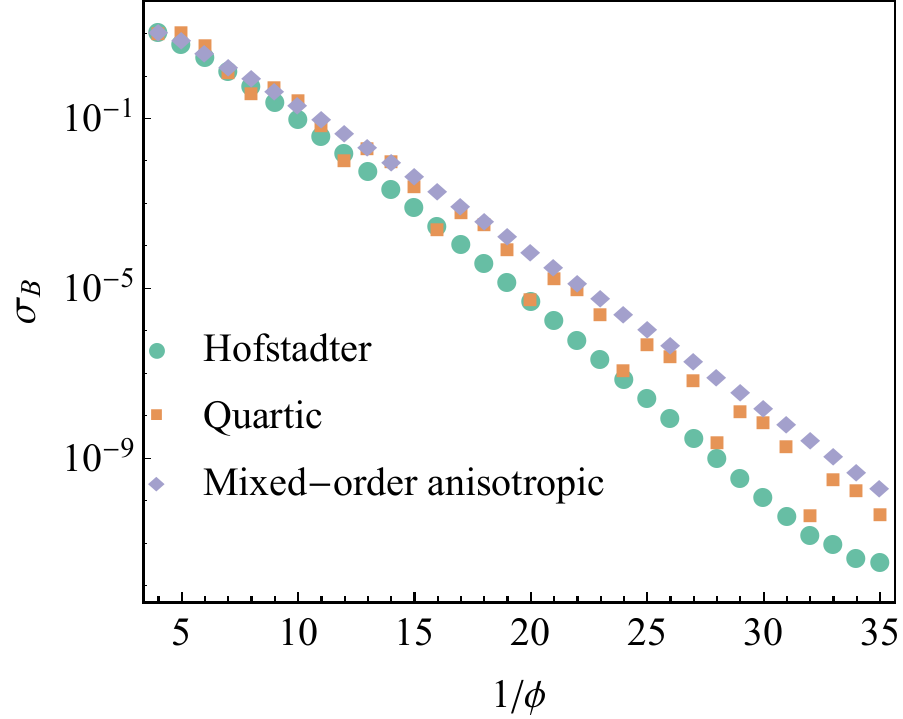}
            \caption{\label{rms-curv-fluct-plot}RMS fluctuation $\sigma_\B$ of Berry curvature $\B$ from its average value on the MBZ. The mixed-order anisotropic model is defined in Eq.~\eqref{eq:mix_aniso}.}
            \end{figure}
            
            To examine the effect of long-range hopping on the tunneling between band minima, we study the ground-state solution of the $n$th-NN Hofstadter model using a WKB approximation for small flux densities, $\phi=1/N$ where $N$ is large~\cite{Watson:1991gy, bender_advanced_1999, Harper:2014vi, Wilmott85}. To start, we consider the Hofstadter Hamiltonian with only $n$th-NN hopping
            \begin{equation}
            H_{n\mathrm{NN}}=-t\sum_{\expval{i,j}_n} a^{\dagger}_i a_j \exp\left\{\mathrm{i}\int_i^j \mathbf{A}\cdot\mathbf{dl}\right\}+\mathrm{H.c.}, 
            \end{equation}
            where $t=1$ is the hopping amplitude, $\mathbf{A}$ is the vector potential, and $\mathbf{dl}$ is an infinitesimal line element from lattice site $i$ to $j$. Without loss of generality, we work in the Landau gauge $\mathbf{A}=-Bx\mathbf{e}_2$, which yields Bloch wavefunctions in the $y$-direction $\psi(x, y)= e^{\mathrm{i}k_y y}\psi(x)$. In the $x$-direction, the states are then characterized by the solutions to the Harper equation~\cite{harper_general_1955}
            \begin{equation}
            -\psi_{x-n} - \psi_{x+n} -2 \cos(n(\phi x - k_y))\psi_x = E \psi_x,
            \end{equation}
            where we have set the lattice spacing $a=1$. In the continuum approximation, this problem may be thought of as a particle moving in a cosine potential. In fact, in the limit of small flux densities, we can take $\psi$ as a continuous function, and approximate the second derivative, such that
            \begin{equation}
            \label{eq:harper}
            -n^2\frac{\psi''(x)}{2} - \cos\left(n\left(\frac{x}{N} - k_y\right)\right)\psi(x) = \frac{(E+2)}{2} \psi(x).
            \end{equation}
            Semi-classically, we therefore expect that the ground-state wavefunction in the $x$-direction consists of Gaussian wavepackets in the troughs of this potential connected weakly via tunneling. Since the derivative of the wavefunction varies slowly for small flux densities, we may solve this differential equation using the WKB method~\cite{Watson:1991gy}. Defining $x'=x/N$ and considering $k_y=0$ without loss of generality, yields the difference equation
            \begin{align}
            \label{eq:harper2}
            \psi(x' + n\phi) + \psi(x' - n\phi) = 2\cosh q(x')\,\psi(x'),
            \end{align}
            for the classically-forbidden region, where $\cosh q(x) = -E/2 - \cos(n x')$. Now substituting the corresponding WKB ansatz~\cite{bender_advanced_1999}
            \begin{align}
            \label{wkb-ansatz}
            \psi(x') = \exp\left[\frac{1}{\phi}\left(S_0(x') + \phi S_1(x')\right)\right],
            \end{align}
            into Eq.~\eqref{eq:harper2}, comparing orders in $\phi$, and solving for $S_0$ and $S_1$ yields
            \begin{align}
            \psi_{\text{exp}}^{\pm}(x') = \frac{1}{\sqrt{\sinh(q(x'))}}\exp\left(\pm \frac{1}{n\phi} \int^x ds\, q(s) \right),
            \end{align}
            for the exponential part of the wavefunction. For further details on the WKB approximation applied to the Harper equation, we refer the reader to the paper by Watson~\cite{Watson:1991gy}. 
            
            In this simple case, the main effect of the longer-range hoppings is to modify the argument of the exponential by a factor $O(1)$ in $\phi$. We expect but do not prove that this is the generic behavior of such solutions. For each of the cases we consider in this article, we observe an overall exponential decay of the amplitudes of fluctuations on the MBZ in numerics. We plot examples of this exponential decay in Fig.~\ref{rms-curv-fluct-plot}.
            
            For this reason, we neglect non-perturbative fluctuations. Since we measure deviations of our Chern bands from Landau levels by band geometric quantities, and since we treat the Berry curvature and FS metric as uniform on the MBZ, deviations from Landau level behavior are quantified using the saturation measures in Eq.~\eqref{geometry-inequalities}.

        \subsection{Hamiltonians with no \texorpdfstring{$h_0$}{h0} term}
        \label{subsec:no_h0}
        
            The fact that the overlap of the ground state of the effective Hamiltonian with the lowest Landau level is large is not surprising, given the large weight of the Landau Hamiltonian $h_0$ in Eq.~\eqref{effective-fock-Hamiltonian}. We might ask whether it is possible to eliminate this term entirely, and what impact this has on the spectrum. In fact, we can isolate the $a^4 + a^{\dag 4}$ term by choosing $t_{01} = t_{10} = 1$, $t_{02} = t_{20} = \frac{1}{8}$, and $t_{11} = - \frac{3}{4}$, giving
            \begin{align}
            H_{\text{eff}} = \frac{3}{2} + \frac{\phi^2}{4}(a^4 + a^{\dag\,4}).
            \end{align}
            This case is distinct from the cases we have considered so far because we are expanding about a local maximum instead of a minimum in the dispersion. The spectrum of this Hamiltonian is not bounded from below, so we cannot find its ground states. However, there is a set of degenerate zero-energy states. While we were unable to obtain an analytical form for the coefficients in the LL basis for the ground state of \eqref{effective-fock-Hamiltonian}, the fact that the states here have exactly zero energy makes the current problem tractable. We find that there are four zero-energy states obtained by unitary transformations of $\ket{0}$, $\ket{1}$, $\ket{2}$, and $\ket{3}$, and we present an analytical calculation of one of the zero-energy states in Appendix~\ref{zero-enegy-quartic}. Although this Hamiltonian does not depend on $h_0$, the state $|\widetilde{0}\rangle = U\ket{0}$ maintains a large overlap $\langle\widetilde{0}|0\rangle \approx 0.987926$ with the Landau level ground state.

        \subsection{Anisotropic models}
        \label{subsec:aniso}
        
            Thus far we have only considered models invariant under $C_4$ symmetry. Of course, generic models will not maintain this symmetry, so we should consider the effects of anisotropy. Rather than treating fully generic models, we consider two straightforward generalizations of the above models that capture essential features. First, we have models in which $K_x$ and $K_y$ both enter the effective Hamiltonian to lowest order in $\phi$, but with different coefficients. The simplest example of such a model is the Hofstadter model with different coefficients in the $x$ and $y$ directions,
            \begin{align}
            \label{aniso-tightbinding}
            H_{\text{TB}} = \alpha(T_1 + T_1^{\dag}) +  \frac{1}{\alpha}(T_2 + T_2^{\dag})
            \end{align}
            with
            \begin{align}
            H_{\text{eff}} = \frac{\alpha}{2} K_x^{2} +  \frac{1}{2\alpha}K_y^{2}.
            \end{align}
            This anisotropic Landau level Hamiltonian and its bands have been studied in relation to nematic QH phases~\cite{maciejko_2013_field}.
            
            Separately, we also consider the case in which we eliminate terms in the effective Hamiltonian at different orders in $\phi$. That is, we choose the hoppings so that
            \begin{align}
            H_{\text{eff}} = h_1 K_x^{2m} + h_2 K_y^{2n}.
            \end{align}
            For example, with $t_{20} = 0$ and $t_{02}=-t_{01}/4$, we have
            \begin{align}
            H_{\text{eff}} = t_{10}&\left(-4 + K_x^2 - \frac{1}{12}K_x^4\right)\nonumber\\
            +t_{01}&\left(-3+\frac{1}{4}K_y^4\right),
            \end{align}
            or, with $t_{01}=t_{10}=1$, 
            \begin{align}
            \label{eq:mix_aniso}
            H_{\text{eff}} = -7 + K_x^2 - \frac{1}{12}K_x^4 +\frac{1}{4}K_y^4.
            \end{align}
            For small $\phi$, the effective Hamiltonian will be dominated by the quadratic term, and we might expect that the problem becomes effectively one-dimensional as $\phi \rightarrow 0$. Calculating the Chern number $C_1$ for the lowest band of the lattice model, we find that $|C_1| = 1$ even for $\phi$ as small as $1/1350$.

    \section{Spatially-inhomogeneous electromagnetic Response}
    \label{sec-em-response}
    
        In this section, we discuss the current response to spatially-inhomogeneous electric fields for QH and Chern band systems, and calculate this response for the families of non-Landau bands introduced in Sec.~\ref{section-single-particle}.
        
        The perturbative response of QH fluids to static, spatially homogeneous electric fields is a key phenomenological feature of such fluids. This response is measured by the conductivity tensor $\sigma$ relating the electric current density $\bm{J}$ and perturbative electric field $\bm{E}$,
        \begin{align}
        \label{conductivity-eq}
        J_{a} = \sigma_{ab}E_b.
        \end{align}
        In particular, one is interested in the transverse, or Hall, conductivity $\sigma_{xy}$, which takes the universal value
        \begin{align}
        \sigma_{xy} = \sigma_{H} = \frac{\nu e^2}{2\pi \hbar}
        \end{align}
        in QH fluids at filling fraction $\nu$.
        
        The FQH fluid carries an emergent geometrical degree of freedom~\cite{haldane_geometrical_2011} corresponding to the shape of elementary composite particles or ``droplets''~\cite{johri_probing_2016}. The geometry of the FQH fluid manifests in part through universal contributions to linear responses corresponding to perturbative variations in geometry, of which the Hall viscosity, $\eta_H$, is an example~\cite{avron_viscosity_1995, tokatly_lorentz_2007, read_non-abelian_2009, haldane_hall_2009}. In topological fluids, the Hall viscosity has a universal part proportional to the topological spin of the fluid~\cite{read_non-abelian_2009}, which is related to the composite droplet shape~\cite{johri_probing_2016}. This geometrical degree of freedom may be obscured by the introduction of non-generic symmetries to the FQH problem implicitly, via the assumption that the underlying single-particle bands are Landau levels.
        
        Recent work has shown that the perturbative response to spatially \emph{inhomogeneous} electric fields is related to the Hall viscosity~\cite{hoyos_hall_2012,bradlyn_kubo_2012}. The universal contribution to the Hall viscosity in a QH fluid on a surface with zero scalar curvature is
        \begin{align}
        \eta_H = \tfrac{1}{2}\hbar\bar{s}\rho_0,
        \end{align}
        where $\rho_0$ is the mean density and $\bar{s}$ is the mean orbital angular momentum or topological spin of the fluid~\cite{read_non-abelian_2009}. For example, a Laughlin fluid with $\nu = \frac{1}{2z+1}$ has,
        \begin{align}
        \bar{s} = z + \tfrac{1}{2}.
        \end{align}
        If we consider the finite-wavevector version of the conductivity tensor relation
        \begin{align}
        J_{a}(\bm{q}) = \sigma_{ab}(\bm{q})E_b(\bm{q}),
        \end{align}
        then the transverse conductivity is
        \begin{align}
        \label{eq:LL_cond}
        \frac{\sigma_{xy}(\bm{q})}{\sigma_H} = 1 + (q\ell)^2\left(\frac{\eta_H}{\hbar \rho_0} - \frac{B^2}{2\nu u_{0}}\pdv[2]{u}{B}\right),
        \end{align}
        where $u(B)$ is the energy density as a function of magnetic field, and $u_0$ is the lowest Landau level energy density 
        \begin{align}
        u_{0} = \frac{\hbar\omega/2}{2\pi\ell^2}.
        \end{align}
        
        Ref.~\onlinecite{harper_finite-wavevector_2018} presents a derivation of two transverse current responses in Chern bands to an applied inhomogeneous electric field. The first is the current per state or current per orbital, which is the current response of a single filled state within a Chern band,
        \begin{align}
        \label{current-per-orbital}
        \expval{I_{yn}} = -i\sum_{m\neq n} &\mel{n,k}{\comm{y}{H}}{m,k}\frac{\mel{m,k}{V(x)}{n,k}}{E_n-E_m}\nonumber\\ &+ \text{H.c.,}
        \end{align}
        with current operator $I_{y} = -i\comm{y}{H}$. The second response is the real-space current density,
        \begin{align}
        \label{current-density}
        J_{yn}(\bm{r}_0) = \sum_{k,m\neq n} &\mel{n,k}{j_y(\bm{r}_0)}{m,k} \frac{\mel{m,k}{V(x)}{n,k}}{E_n-E_m}\nonumber \\ &+ \text{H.c.,}
        \end{align}
        where $j_y(\bm{r}_0)$ is the symmetrized current density operator
        \begin{align}
        j_y(\bm{r}_0) = \frac{1}{2}\left(I_y \delta(\bm{r} - \bm{r}_0) + \delta(\bm{r} - \bm{r}_0)I_y \right).
        \end{align}
        From the current density response, we can calculate the transverse conductivity $\sigma_{xy}(\bm{q})$. This expression for $J_{yn}(\bm{r}_0)$ is equivalent to that obtained by a linear response calculation~\cite{harper_finite-wavevector_2018}. The factor of $H$ in the current per orbital cancels the energy denominator, so that $\expval{I_{yn}}$ depends only on the band projectors and not on the energy dispersion. In contrast, this cancellation does not occur in the case of the current density. For a generic scalar potential $V(x)$, we can write a Taylor series expansion about $x_0$,
        \begin{align}
        V(x) = \sum c_p (x - x_0)^p.
        \end{align}
        Then the current per orbital in the $n$-th Landau level in response to the potential $V(x)$ is, in terms of the expansion coefficients $c_p$,
        \begin{align}
        \label{current-per-orbital-expansion}
        \expval{I_{yn}} = -c_1\ell^2 &- 3c_3 \ell^4\left(n+\frac{1}{2}\right)\nonumber\\ &- \frac{15}{2}c_5\ell^6 \left(n^2 + n + \frac{1}{2}\right) + \ldots,
        \end{align}
        and the current density is
        \begin{align}
        \label{current-density-expansion}
        J_{yn}(\bm{r}_0)=-\frac{1}{2\pi\ell^2}[&c_1\ell^2 + 9c_3 \ell^4\left(n+\frac{1}{2}\right) \nonumber\\ + &\frac{5}{2}c_5 \ell^6\left(11 + 30n + 30n^2\right) + \ldots].
        \end{align}
        
        In the case of non-Landau level models, the above expressions \eqref{current-per-orbital} and \eqref{current-density} still hold, with the caveats that we must calculate the current operator $-i[y,H]$ using the non-Landau Hamiltonian and interpret the states $\ket{n,k}$ as eigenstates of this Hamiltonian. We have seen that we can obtain these perturbative eigenstates numerically by considering a finite-dimensional truncation of the Hamiltonian written in the Landau level basis. In this way, we can numerically calculate these current responses for the case of our non-Landau bands. This calculation is carried out in detail for $C_4$-symmetric models in Ref.~\onlinecite{harper_finite-wavevector_2018}.
        
        Here, we compute the corrections to the Landau level responses \eqref{current-per-orbital-expansion} and \eqref{current-density-expansion} for the Hamiltonians
        \begin{align}
        \label{eq:H2n}
        H^{(2n)} = K_x^{2n} + K_y^{2n}.
        \end{align}
        Writing $\ket{\lambda^{(2n)}}$ for the eigenstates of this Hamiltonian, we obtain the current per orbital
        \begin{align}
        \expval{I_{y\lambda}}^{(2n)} = 2\sum_{\substack{\mu\neq\lambda\\p}}&(-1)^pc_p\,\ell^{2p} \mel{\lambda}{K_y^{2n-1}}{\mu}\frac{\mel{\mu}{K_y^p}{\lambda}}{E_{\lambda}-E_{\mu}} + \text{H.c.}
        \end{align}
        and current density
        \begin{align}
        \expval{J_{y\lambda}}^{(2n)} = \frac{1}{\pi \ell^2}\sum_{\substack{\mu\neq\lambda\\p,r}} (-1)^{r} c_p^2\,\ell^{2p} \binom{p}{p-r}\nonumber\\\times\mel{\lambda}{K_y^{2n-1+p-r}}{\mu}\frac{\mel{\mu}{K_y^r}{\lambda}}{E_{\lambda}-E_{\mu}} + \text{H.c.}.
        \end{align}
        
        \begin{figure}[tb]
        \includegraphics[width=3.0in]{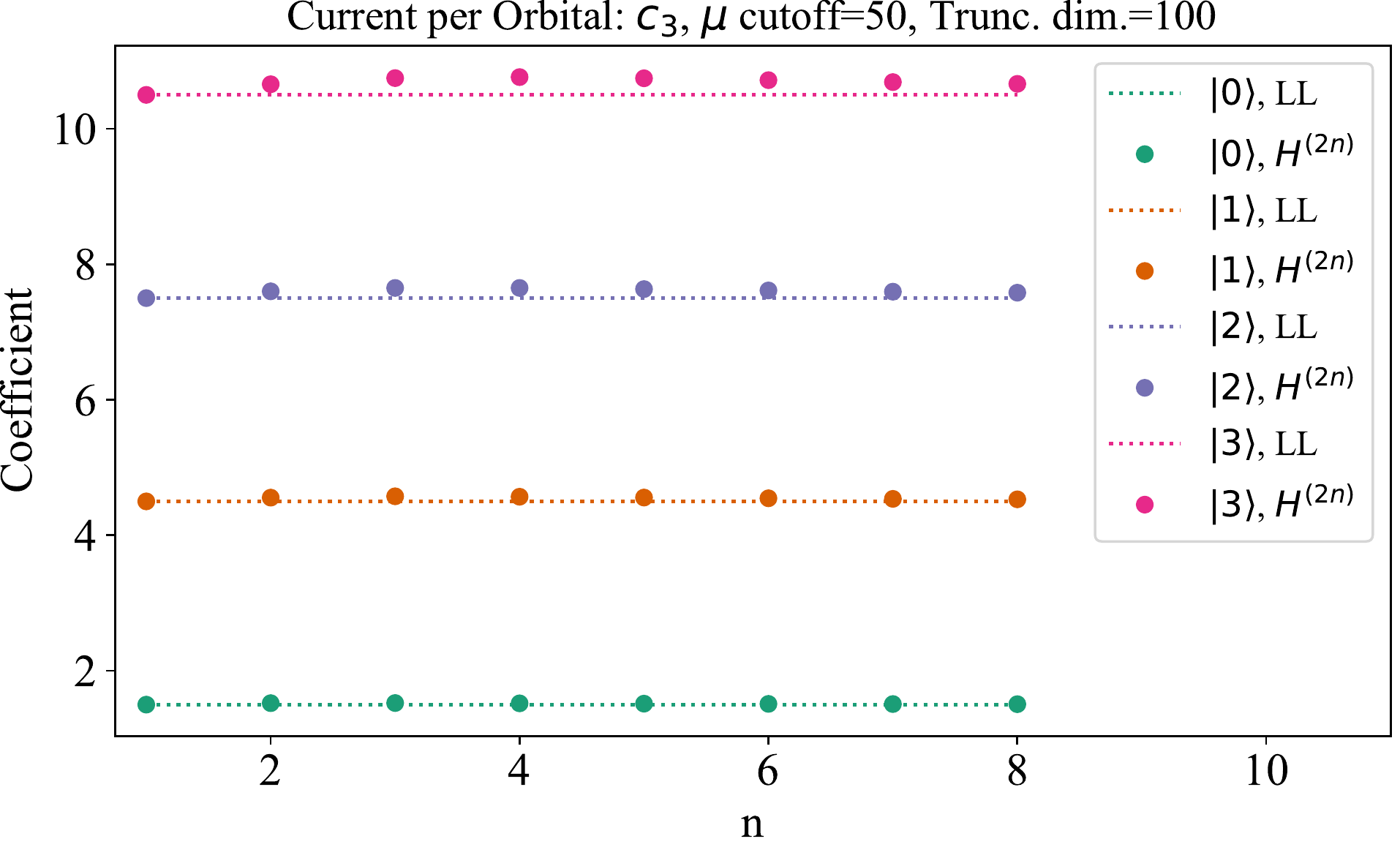}
        \includegraphics[width=3.0in]{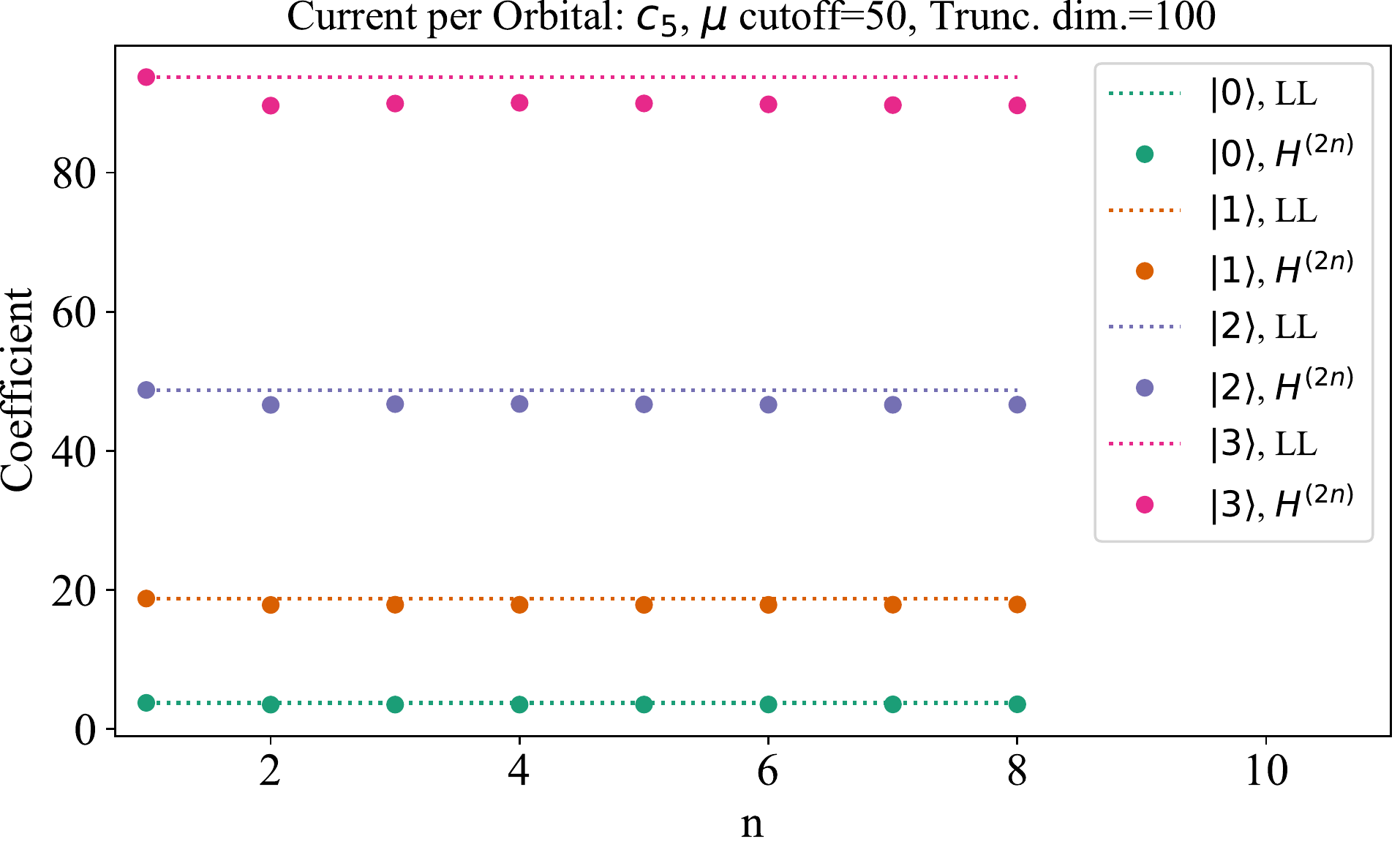}
        \caption{\label{plot-current-per-orbital} Summary of current per orbital response of the bands of $H^{(2n)}$ to inhomogeneous electric potential $V(x) = \sum_p c_p x^p$. For each value of $n$, we plot the numerical factor multiplying $c_3$ (top) and $c_5$ (bottom) in the current per orbital \eqref{current-per-orbital-expansion}. The dotted lines show the values of these factors for the corresponding Landau levels.}
        \end{figure}
        
        \begin{figure}[tb]
        \includegraphics[width=3.0in]{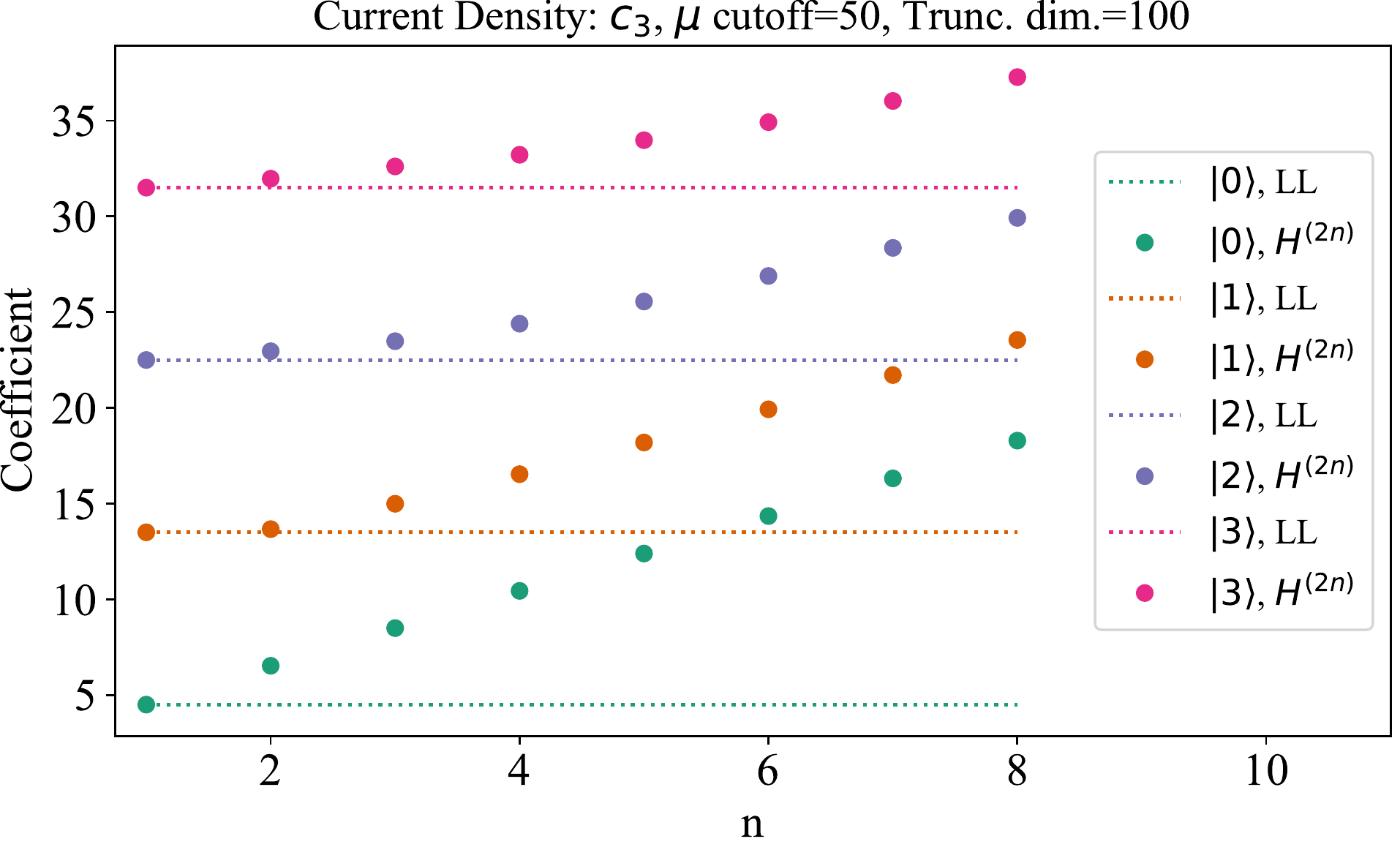}
        \includegraphics[width=3.0in]{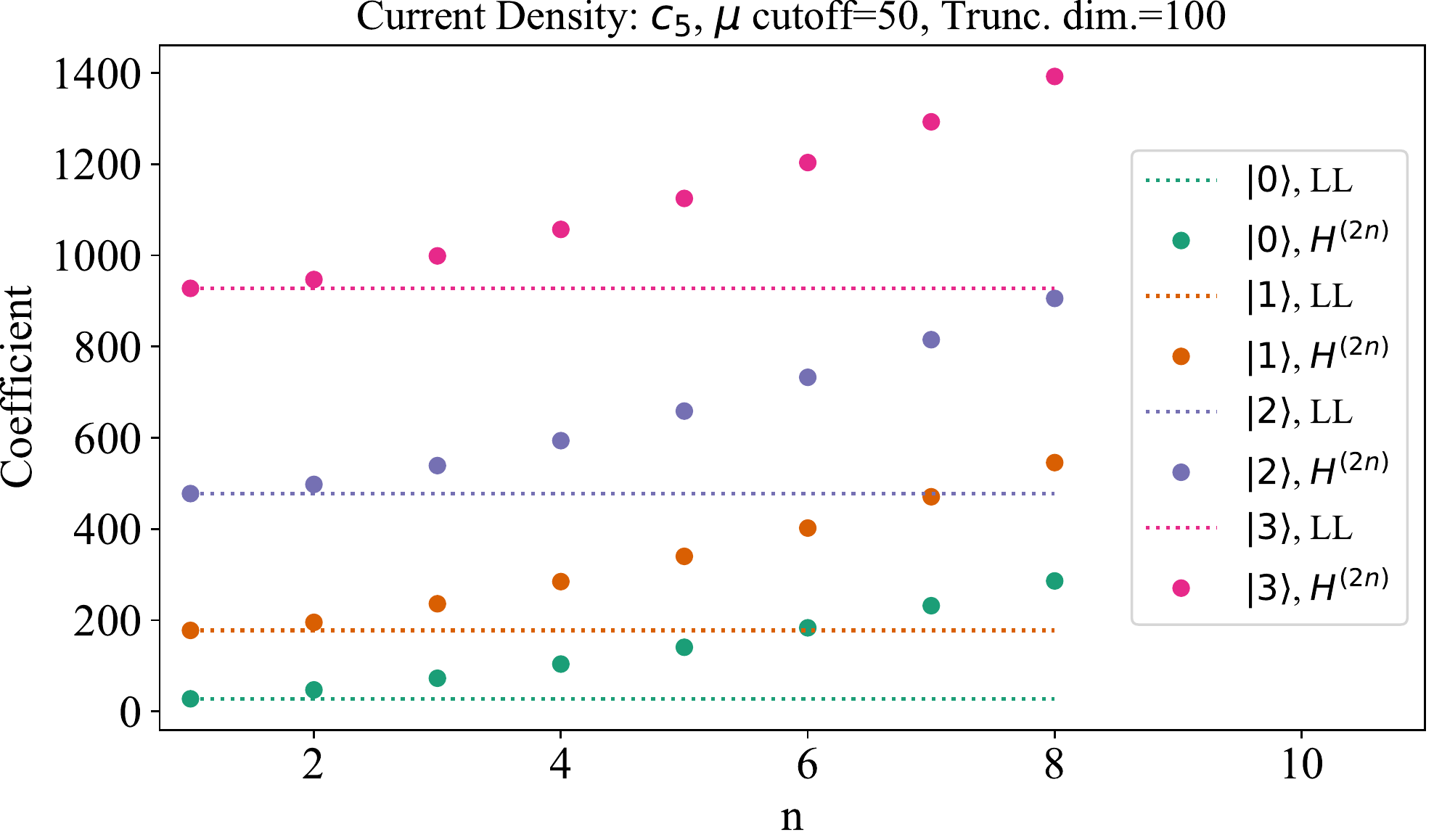}
        \caption{\label{plot-current-density}Summary of current density response of the bands of $H^{(2n)}$ to inhomogeneous electric potential $V(x) = \sum_p c_p x^p$. For each value of $n$, we plot the numerical factor multiplying $c_3$ (top) and $c_5$ (bottom) in the current density \eqref{current-density-expansion}. The dotted lines show the values of these factors for the corresponding Landau levels.}
        \end{figure}
        
        The effect of these corrections amounts to employing the expressions \eqref{current-per-orbital-expansion} and \eqref{current-density-expansion} with changes to the numerical factors multiplying each expansion coefficient $c_p$. For example, in the Landau level case, the numerical factors multiplying $c_1$, $c_3$, and $c_5$ in Eq.~\eqref{current-per-orbital-expansion} are respectively 1, 3/2, and 15/4 for $n=0$, and 1, 9/2, 75/4 for $n=1$. In the non-Landau level case, the factor multiplying the coefficient $c_1$, which is the spatially-homogeneous component of the electric field corresponding to the Hall conductivity, does not change from its Landau level value. In Fig.~\ref{plot-current-per-orbital}, we plot the numerical factors multiplying $c_3$ and $c_5$ in the current per orbital for some eigenstates $\ket{\lambda}$ of $H^{(2n)}$, and in Fig.~\ref{plot-current-density}, we plot the same for the current density. The coefficients for the current per orbital response do not diverge significantly from their Landau level values compared to the coefficients for the current density response, which reflects the current density's dependence on the energy dispersion. We note that the data presented are insufficient to comment on the current per orbital response deviation with respect to $n$, or the relative electromagnetic response deviation with respect to state number. For the current density, the coefficient $c_3$ corresponds to the Hall viscosity in the Landau level case, as seen by comparing Eqs.~\eqref{eq:LL_cond} and~\eqref{current-density-expansion}. In the case of lattice models, the notion of a Hall viscosity is more difficult to define~\cite{haldane_geometry_2015}. However, $c_3$ can be readily derived for the effective continuum Hamiltonian in Eq.~\eqref{eq:H2n} and matched with the data in Fig.~\ref{plot-current-density}.
        
    \section{Localization-delocalization transition}
    \label{sec-localization}
    
        In this section, we review the theory of the localization-delocalization transition in QH systems in Sec.~\ref{subsec:background_theory}, we outline the transfer matrix method for computing the localization length exponent in Sec.~\ref{subsec:transfer_matrix}, and we present our result for the zero-quadratic model in Sec.~\ref{subsec:loc_length}.
        
        \subsection{Background theory}
        \label{subsec:background_theory}
    
		The localization length, $\xi$, of an electronic wavefunction arises in the phenomena of Anderson localization wherein wavefunctions are exponentially localized around some position, $\textbf{m}_0$, by a random disorder potential~\cite{anderson1958absence,lee1985disordered}
		\begin{align}
		|\psi(\textbf{m})| \propto \exp(-|\textbf{m}-\textbf{m}_0|/\xi).
		\end{align}
		
		In the integer quantum Hall effect (IQHE), electronic states are localized at all energies except the Landau level energies, $E_n$, where the electronic states become extended.
		Near the critical energies, $E_n$, the localization length varies with critical exponent $\nu$ as
		\begin{align}
		\xi(E) \propto |E-E_n|^{-\nu}.
		\end{align}
		
		The value of this critical exponent has drawn much attention, with experimental values of approximately $\nu=2.38$~\cite{wei1988experiments,li2009scaling}. Theoretical values are less clear, with recent values between $2.37$ and $2.62$ being reported for different models~\cite{slevin2009critical,obuse2012finite,gruzberg2017geometrically,klumper2019random,lutken2007geometric,lutken2019elliptic,puschmann2019integer,ippoliti2018integer,zhu2019localization,sbierski2020criticality,huang2020numerical,puschmann2021green}. A recent large-scale numerical study of the disordered Hofstadter model using the recursive Green's function method found $\nu=2.58(3)$~\cite{puschmann2019integer}.
		
		The disordered Hofstadter model is a lattice version of the IQHE in the limit of small $\phi$~\cite{hofstadter_energy_1976,huckestein1995scaling,puschmann2019integer}.
		Landau level energies vary as $E_n(\phi)=\hbar\omega_B(n+\tfrac{1}{2})=(\hbar^2/m^*a^2)(2\pi\phi/\phi_0)(n+\tfrac{1}{2})$, while for the Hofstadter model at small $\phi$,
		\begin{align}
		E_n(\phi) = -4 + 2(2\pi\phi/\phi_0) (n+\tfrac{1}{2}),
		\end{align}
		which is the same as the Landau level energies shifted by $4$ if the effective mass is taken to be $m^*=\hbar^2/2a^2$.
		
		In contrast, in the zero-quadratic model, the level energies have a quadratic dependence on $\phi$ at small $\phi$,
		\begin{align}\label{eq:quadratic-energies}
		E_n(\phi) = -3 + \frac{(2\pi\phi/\phi_0)^2}{8}\left(6(n+\tfrac{1}{2})^2 + \frac{3}{2}\right).
		\end{align}
		We can understand this energy dependence as \eqref{effective-fock-Hamiltonian}, but modified by removing the contribution of $a^4+a^{\dagger 4}$, which disappears for small flux. To quantify deviations from the quadratic behavior at finite flux, we note that at $\phi/\phi_0<1/100$, the energies of \eqref{eq:quadratic-energies} are all within 10\% of the true values that account for the $a^4+a^{\dagger 4}$ term.
		
		We then ask: does the zero-quadratic model at small $\phi$ exhibit the same critical behavior in its electronic localization-delocalization transition as in Landau levels as evidenced by the electronic localization-length critical exponent? To answer this question, we use the transfer matrix method and a finite-size scaling analysis. Doing so, we find a critical exponent of $\nu=2.57(3)$, which is in good agreement with the range of proposed critical exponents for the IQHE, but larger than the values measured in experiments~\cite{wei1988experiments,li2009scaling}.
		
		\subsection{Transfer Matrix Method}
		\label{subsec:transfer_matrix}
			
			Here, we derive a transfer matrix for a finite-size system, and determine the localization length from the set of transfer matrices. We consider a cylindrical geometry with width $W$ sites with periodic boundary conditions and length $L$ sites.
			The transfer matrix $A_\ell$, takes the wavefunction $|\ell\rangle \in \mathbb{C}^W$ for a layer $\ell$ and transfers it to the wavefunction of the next layer $|\ell+1\rangle$,
			\begin{align}
			\begin{pmatrix}
			\ell+2\\\ell+1\\\ell\\\ell-1
			\end{pmatrix}
			= A_\ell \begin{pmatrix}
			\ell+1\\\ell\\\ell-1\\\ell-2
			\end{pmatrix}.
			\end{align}
			Evidently $A_\ell$ acts on the layer wavefunctions as
			\begin{align}
			\begin{pmatrix}
			\ell+2\\\ell+1\\\ell\\\ell-1
			\end{pmatrix}
			= \begin{pmatrix}
			\square&\square&\square&\square\\
			1&0&0&0\\
			0&1&0&0\\
			0&0&1&0
			\end{pmatrix} \begin{pmatrix}
			\ell+1\\\ell\\\ell-1\\\ell-2
			\end{pmatrix},
			\end{align}
			for some as-yet undetermined matrices, $\square$. Here, $1$ is the $W\times W$ identity matrix, and $0$ is the $W\times W$ zero matrix.
			
			Now, the time-independent Schr\"odinger equation is $H|\ell\rangle = E|\ell\rangle$, so with $H=H_\mathrm{TB}+V$ for an onsite quenched disorder potential given by $V=\sum_{\bm{m}} \epsilon_{\bm{m}} \create{\bm{m}}\annihilate{\bm{m}}$, we have
			\begin{align}
			-t_2 |\ell\!+\!2\rangle -t_1|\ell\!+\!1\rangle + \bar{H}_\ell |\ell\rangle -t_1|\ell\!-\!1\rangle -t_2 |\ell\!-\!2\rangle = E|\ell\rangle,
			\end{align}
			where the intra-layer Hamiltonian is, with $\bm{m}=(\ell,w)$,
			\begin{align}
			\bar{H}_\ell = \big(\sum_{w} \epsilon_{\ell,w} \create{\ell,w}\annihilate{\ell,w}\big)
			- t_1(T_{y,\ell}+T^{\dagger}_{y,\ell})
			- t_2(T_{y,\ell}^{2}+T^{\dagger 2}_{y,\ell}),
			\end{align}
			with the intra-layer translation operators
			\begin{align}
			T_{y,\ell} = \sum_w e^{-i(2\pi\phi/\phi_0)\ell} \create{\ell,w+1} \annihilate{\ell,w}.
			\end{align}
			
			Rearranging, we find that
			\begin{align}
			A_\ell =
			\begin{pmatrix}
			(t_1/t_2)1 & (\bar{H}_\ell-E1)/t_2 & (t_1/t_2)1 & 1\\
			1 & 0 & 0 & 0\\
			0 & 1 & 0 & 0\\
			0 & 0 & 1 & 0
			\end{pmatrix}.
			\end{align}
			Thus, the component of the wavefunction in the last layer is given in terms of the wavefunction in the first layer as
			\begin{align}
			|L\rangle = \left(\prod_{\ell=1}^{L} A_\ell\right) |1\rangle
			= \left(Q_L \prod_{\ell=1}^L R_\ell\right) |1\rangle,
			\end{align}
			where the product of transfer matrices was decomposed into an orthogonal matrix $Q_L$, and upper-triangular matrices $R_\ell$ using the $QR$ decomposition~\cite{pichard1981finite,mackinnon1981one,mackinnon1983scaling,kramer1993localization}.
			
			Now, the localization length is related to the transfer matrix through the set of Lyapunov exponents $\gamma_w$~\cite{pichard1981finite,mackinnon1981one,mackinnon1983scaling,kramer1993localization}, such that
			\begin{align}
			\xi = \frac{1}{\min_w|\gamma_w|},
			\end{align}
			where the Lyapunov exponents are given by the sum of the eigenvalues of $A_\ell$ (diagonal elements of $R_\ell$) scaled by the logarithm,
			\begin{align}
			\gamma_w = \frac{1}{L} \sum_{\ell=1}^L \ln|R_{\ell}^{(w,w)}|.
			\end{align}
		
		\subsection{Localization-Length Exponent}
		\label{subsec:loc_length}
			
			We determined the localization length as a function of energy and strip width for the lowest level of the zero-quadratic model at $\phi=(1/10)(\phi_0/2\pi)$ for which the energy-dependence on the field is very close to quadratic and the magnetic length is still fairly small, with strip widths of $12,16,24,32,48,64,96,$ and $128$ sites with periodic boundary conditions. 51 energies were selected linearly spaced in the neighborhood of the critical energy $E_0((1/10)(\phi_0/2\pi))=-2.88920t_1$. For the quenched disorder potential, we used uncorrelated white noise taken from a uniform distribution, $V_{\ell,w}=\mathrm{uniform}(-s/2,s/2)$ with disorder strength $s=0.1t_1$, a value which is large compared to the intrinsic level width of $1.1\times10^{-4}t_1$, but smaller than the level spacing $0.400t_1$. We did this for 20 realizations of strips of length $L=10^5$, for a cumulative effective strip length of $L=2\times10^6$.
			
			\begin{figure}[thp]
				\includegraphics[width=1.1\linewidth]{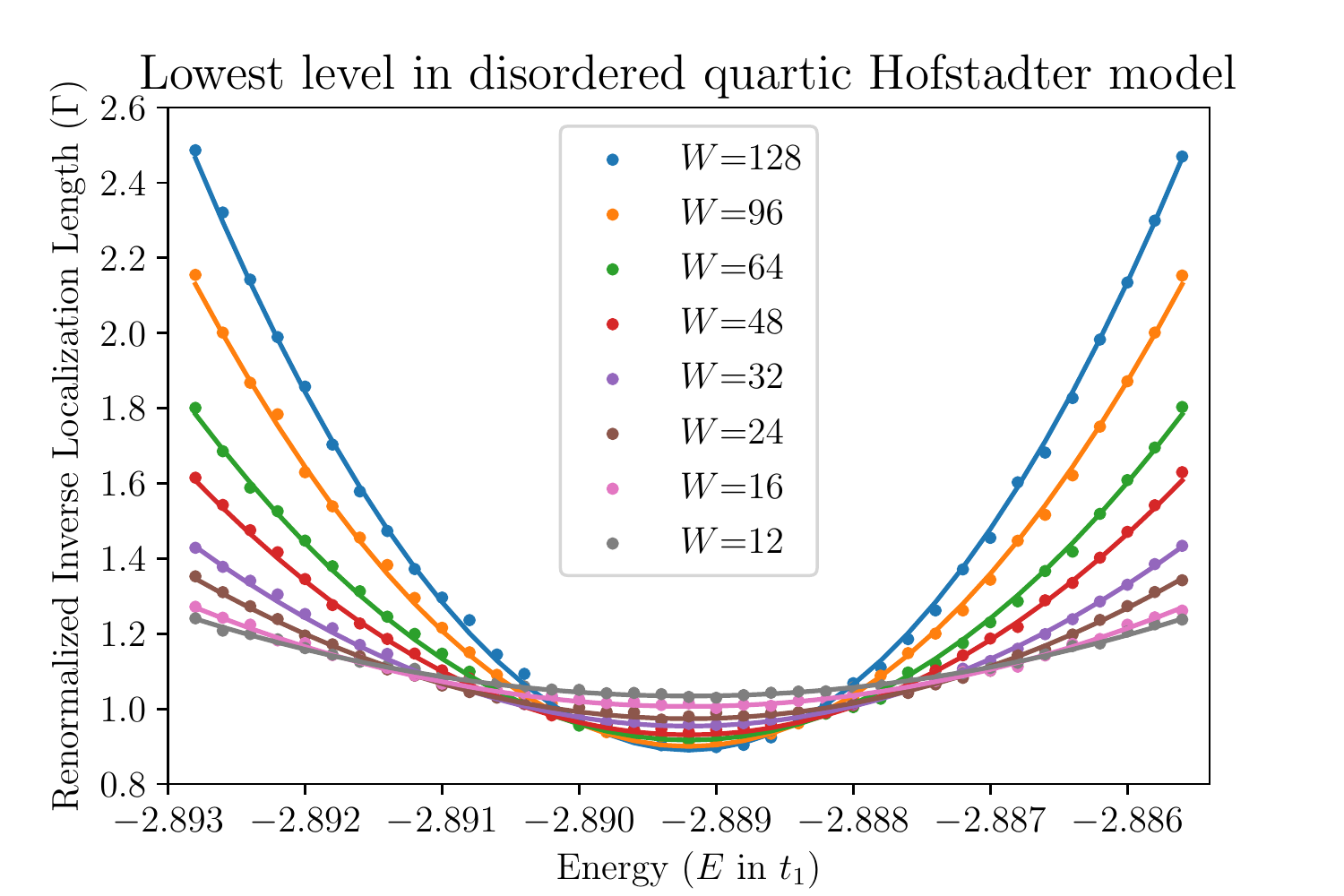}
				\caption{\label{fig:localization} Renormalized inverse localization length in the disordered zero-quadratic model at flux $\phi=(1/10)(\phi_0/2\pi)$ and disorder strength $s=0.1t_1$ near the center of the lowest level. Points: renormalized inverse localization-lengths as a function of width. Lines: fit to $\Gamma(E,W)$ using Eq.~\eqref{eq:fit} with parameters taken as the average of the values determined from the fits to 100 synthetic data sets $\widetilde{\Gamma}(E,W)=\Gamma(E,W)+\mathrm{gaussian}(0,\sigma(E,W))$, where $\sigma(E,W)$ is the standard deviation of the data point $\Gamma(E,W)$ across the $20$ realizations. The mean fit value for $\nu$ across the $1000$ data sets is $\nu=2.57$ with a standard deviation of $0.03$.}
			\end{figure}
			
			We then used a finite-size scaling analysis to analyze the data~\cite{slevin2009critical,obuse2012finite,puschmann2019integer,zhu2019localization}. First, we calculated the renormalized inverse localization lengths $\Gamma = W/\xi$ from the mean of the localization lengths of all the realizations, and determined the uncertainty in the mean by comparing the localization lengths for the different realizations. We then performed a least-squares regression with the function:
			\begin{align}\label{eq:fit}
			\Gamma(E,W) = a_0 + a_2(E-E_0)^2W^{2/\nu} + b_1 W^{-y}.
			\end{align}
			Here, $a_0$, $a_2$, $b_1$, $\nu$, and $y$ are fit parameters. This function is the lowest-order non-trivial symmetric function with an irrelevant scaling term~\cite{obuse2012finite}. To understand the uncertainty in the fit of the exponent, we generated $100$ synthetic data sets where each data point was the mean of the realizations plus a normally distributed random number with zero mean and standard deviation equal to the standard deviation of the data point across the $20$ realizations. We then performed a least-squares fit of the above function to the data with each of the synthetic data sets, as shown in Fig.~\ref{fig:localization}. The exponents determined for the synthetic data sets were normally distributed, where we found: $\nu=2.57(3)$ and $y=0.31(4)$.
			
			This value of $\nu$ is in close agreement with the value determined by Puschmann~et al.~for the disordered Hofstadter model where they found $\nu=2.58(3)$~\cite{puschmann2019integer}. On the one hand, this is unsurprising since the overlap of the states with Landau levels is close to 1, however this finding does show that the quantum geometry of these non-Landau level states does not noticeably affect the critical behavior of the localization-delocalization transition. This is consistent with findings that the localization-delocalization transition is independent of many model details, as exemplified by the success of the Chalker-Coddington model~\cite{chalker1988percolation}. We also find that this result persists as we vary the value of $t_2$ in Eq.~\eqref{quartic-harper}, as discussed in Appendix~\ref{exponentNNN}.

    \section{Band geometry and stability of FQH states}
    \label{sec-geometry}
    
        In the preceding sections, we introduced a series of tight-binding models and studied their electromagnetic response and localization length behavior. When $\phi$ is small, the single-particle eigenstates of these models are effectively mixtures of Landau levels. In this section, we study the stability of FQH phases in these bands. We start our discussion by reviewing the connection between band geometry and FQH stability in Chern bands. To study the FQHE, we introduce a repulsive, two-body interaction potential on the lattice,
        \begin{align}
        V = \sum_{\bm{m},\bm{n}}v_{\bm{m}\bm{n}}\, c^{\dag}_{\bm{m}} c_{\bm{m}} c^{\dag}_{\bm{n}} c_{\bm{n}}.
        \end{align}
        We fractionally fill a single band of energy eigenstates with $N_p$ particles. Each band contains $N_s = A_{\text{tot}}/A_{\text{MUC}}$ states and the filling fraction is $\nu = N_p /N_s$. We work within a single band with projector $\mathcal{P}$; that is, we diagonalize the interaction Hamiltonian $H_{\text{int}} = \mathcal{P}V\mathcal{P}$. The single-mode approximation (SMA) provides one route toward understanding the stability of FQH phases, by estimating the spectrum of collective density excitations above the ground state~\cite{girvin_magneto-roton_nodate} using single-mode density operators. In applying the SMA to FQH states, Girvin-MacDonald-Platzmann (GMP) found that the Landau level-projected density mode operators
        \begin{align}
        \bar{\rho}_{\bm{q}} = e^{-q^2\ell^2}\mathcal{P}\rho_{\bm{q}}\mathcal{P} = e^{-q^2\ell^2}\mathcal{P}e^{i\bm{q}\cdot \bm{r}}\mathcal{P}
        \end{align}
        obey the GMP algebra~\cite{girvin_magneto-roton_nodate}
        \begin{align}
        \comm{\bar{\rho}_{\bm{q}}}{\bar{\rho}_{\bm{q}'}} = 2i\sin\left(\frac{\ell^2}{2}\epsilon_{ab}q_a q'_b\right)\bar{\rho}_{\bm{q} +\bm{q}'}.
        \end{align}
        We have written these expressions for a single particle, but they also hold for the many-particle density operators $\bar{\rho}_{\bm{q}} = e^{-q^2\ell^2}\mathcal{P}\sum_{i}e^{i\bm{q}\cdot\bm{r}_i}\mathcal{P}$. These commutation relations describe the algebra of symmetries of the Landau levels. The significance of the projected density operators to the interaction spectrum of the FQH problem is their appearance in the projected interaction Hamiltonian
        \begin{align}
        H_{\text{int}} = \mathcal{P}V\mathcal{P} = \frac{1}{2}\sum_{\bm{q}} v_{\bm{q}} \bar{\rho}_{\bm{q}}\bar{\rho}_{-\bm{q}}.
        \end{align}
        
        When the density mode operators are projected to a Chern band instead of a Landau level, they do not necessarily form a closed algebra~\cite{parameswaran_fractional_2012}. We observe this by expanding the commutation relation to lowest order in $\bm{q}$:
        \begin{align}
        \comm{\bar{\rho}_{\bm{q}}}{\bar{\rho}_{\bm{q}'}} &= -q_a q'_b \comm{\mathcal{P}r_a\mathcal{P}}{\mathcal{P}r_b\mathcal{P}} + O(\bm{q}^3)\\
        &= - i \epsilon_{ab} q_a q'_b \mathcal{P} \sum_{\bm{k}} \B(\bm{k}) \mathcal{P}_{\bm{k}} + O(\bm{q}^3).
        \end{align}
        If the Berry curvature is uniform, $\B(\bm{k}) = \B_0$, then the relation
        \begin{align}
        \comm{\bar{\rho}_{\bm{q}}}{\bar{\rho}_{\bm{q}'}} = - i \B_0\epsilon_{ab} q_a q'_b \mathcal{P} + O(\bm{q}^3)
        \end{align}
        reproduces the Landau level relation
        \begin{align}
        \comm{\bar{\rho}_{\bm{q}}}{\bar{\rho}_{\bm{q}'}} = i \ell^2 \epsilon_{ab} q_a q'_b \mathcal{P}_{\text{LL}} + O(\bm{q}^3),
        \end{align}
        with the magnitude of the Berry curvature $\B_0$ replacing the magnetic length $\ell^2$ as the unit area. Just as the uniformity of the Berry curvature implies closure of the Chern band GMP algebra to lowest order in $\bm{q}$, additional conditions on the geometry of the Chern band imply closure at higher orders~\cite{roy_band_2014}. In particular, the algebra closes to $O(\bm{q}^3)$ if both the Berry curvature and FS metric are uniform. If both $\mathcal{B}(\bm{k})$ and $g(\bm{k})$ are uniform, then saturation of the determinant equality or equivalently the minimization of the DISM for all $\bm{k}$, $\mathcal{D}(\bm{k}) = 0$, implies closure to all orders in $\bm{q}$.
        
        The role of these geometric quantities in the Chern band GMP algebra motivates the GSH~\cite{roy_band_2014,jackson_geometric_2015}, which states that bands that are geometrically similar to Landau levels are more favorable to stabilize FQH phases. Numerical evidence consistent with the GSH has been observed in FCIs~\cite{jackson_geometric_2015} and Hofstadter systems~\cite{bauer_quantum_2016}. However, these studies leave open questions about the role of geometry in FQH stability. In FCI models, fluctuations in band-geometric densities are generically non-negligible, and are correlated with saturation of the inequalities. By contrast, in the weak-field limit of the Hofstadter model, we may neglect fluctuations, but the lack of tunable model parameters means that we cannot vary the band-geometric inequalities independently from $\phi$. The more generic tight-binding models we study in this work therefore provide a new regime in which to study the role of band geometry in the stability of lattice FQH phases. Considering, for example, the model \eqref{quartic-harper}, we can set $\phi$ to be arbitrarily small and neglect fluctuations while varying $t_2$ and hence $\expval{\mathcal{T}}$.
        
        The projected interaction necessarily depends on the form of the interaction coefficients $v_{\bm{q}}$ in addition to the projected density operators. While the GSH may indicate the favorability of single-particle bands for hosting FQH phases, it is really the interplay between the interaction $v_{\bm{q}}$ and the projected density operators that determines stability. Since the SMA is fundamentally a \emph{long-wavelength} approach to studying the behavior of FQH fluids, we expect that its applicability depends on the existence of effective continuum descriptions of both the interaction and the single-particle bands.
        
        We consider systems of $N_p = 8$ fermions and bosons. For fermions, we use a system of $6 \times 4$ MUCs, so that $\nu=1/3$, each of size $4m \times 6m$ for various integers $m$. This ensures that the overall system size is the same in each direction. For bosons, we use a system of $4 \times 4$ MUCs of size $m \times m$, and $\nu=1/2$. For each set of parameter values, we verify the presence of an excitation gap above a $q$-fold quasidegenerate space of ground states, as expected for a $\nu=1/q$ Laughlin fluid. This provides evidence that the bands of the tight-binding model may host FQH phases even when the hopping amplitudes are tuned to eliminate the quadratic term in the Hamiltonian.
        
        \begin{figure*}[t]
        \includegraphics[width=2.15in]{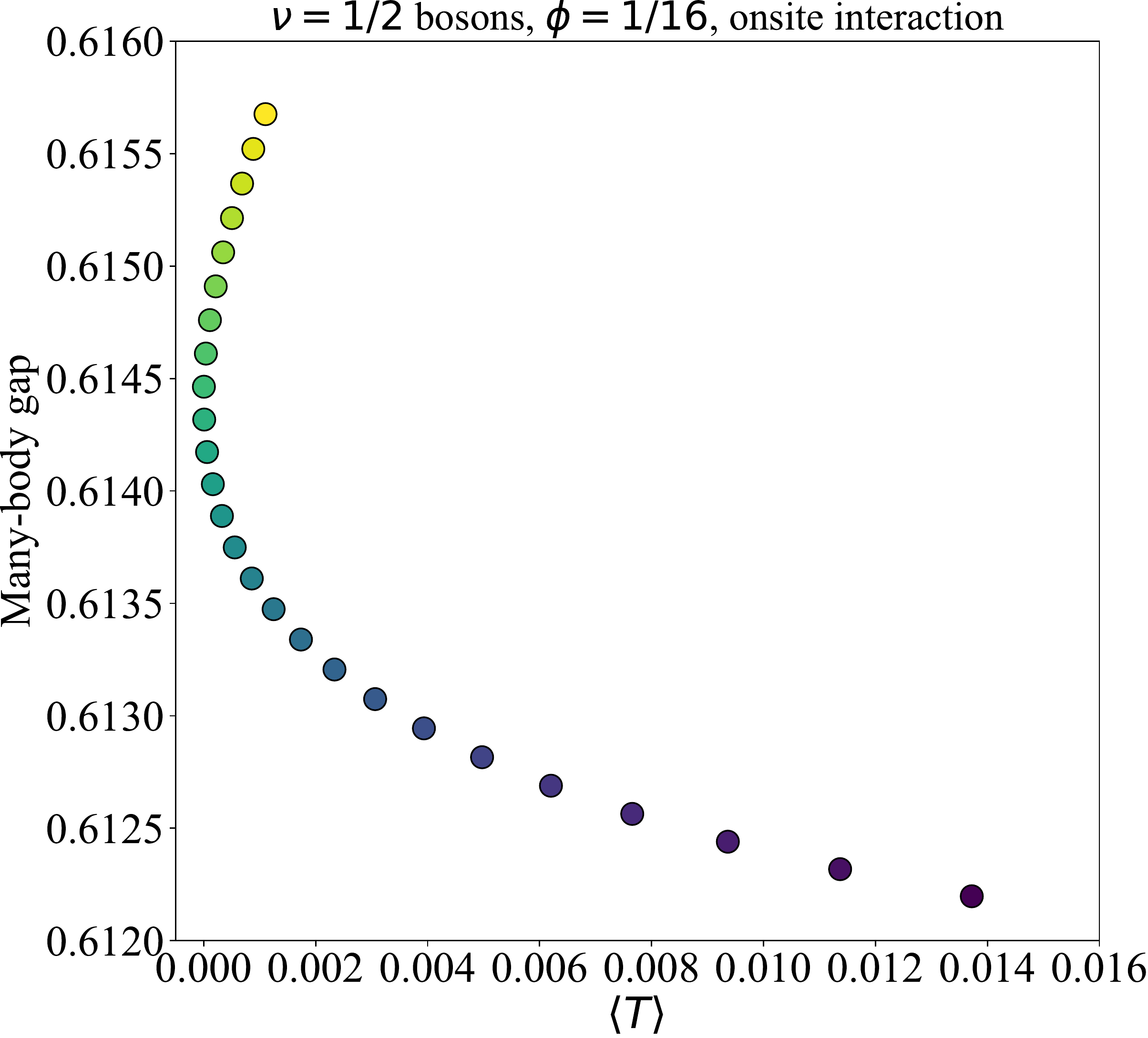}
        \includegraphics[width=2.2in]{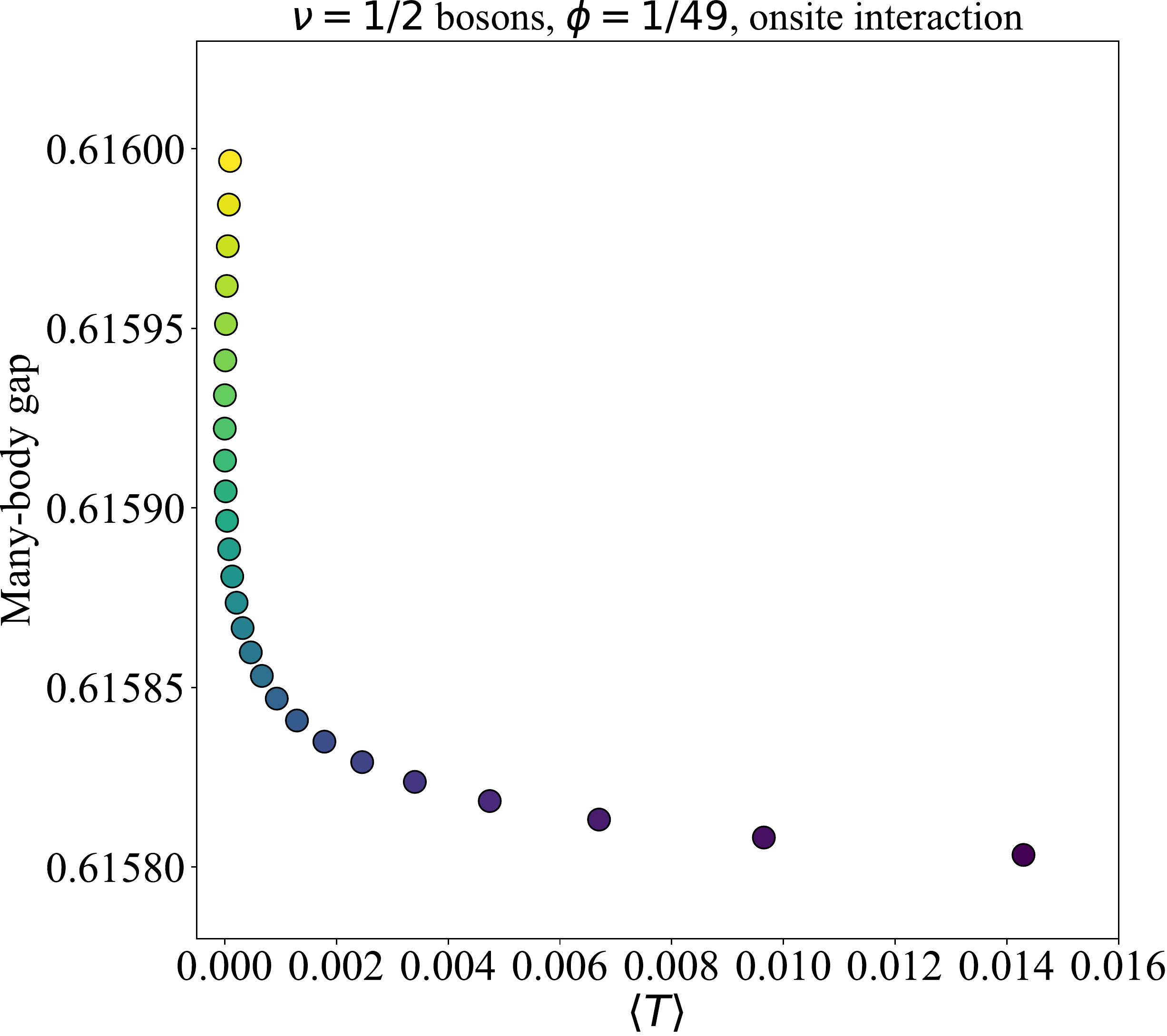}
        \includegraphics[width=2.45in]{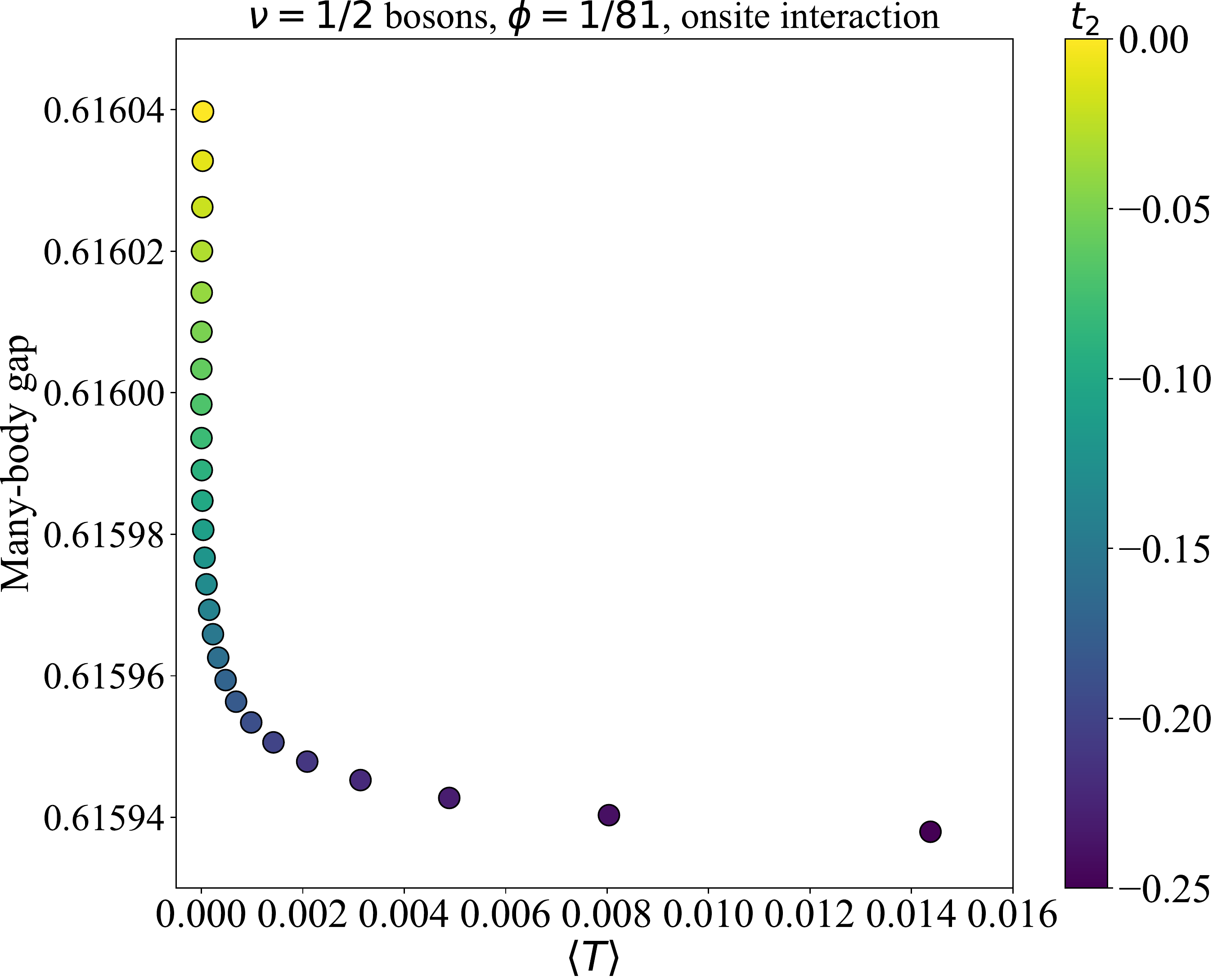}
        \caption{\label{gap-v-trace-bosons}Gap between first excited state and 2-fold degenerate ground state for $N_p = 8$ bosons occupying the ground state band of the tight binding model \eqref{quartic-harper} at filling $\nu=1/2$. The interaction between bosons is an onsite, infinite density-density repulsion. We plot the dependence of the gap on the TISM $\expval{\mathcal{T}}$ as the NNN hopping parameter $t_2$ is varied. Each panel shows a different value of flux per plaquette $\phi$: from left to right we have $\phi=1/16$, $\phi=1/49$ and $\phi=1/81$. For each case, the largest gap is observed for the ordinary Hofstadter model with $t_2 = 0$.}
        \end{figure*}
        \begin{figure*}[t]
        \includegraphics[width=2.25in]{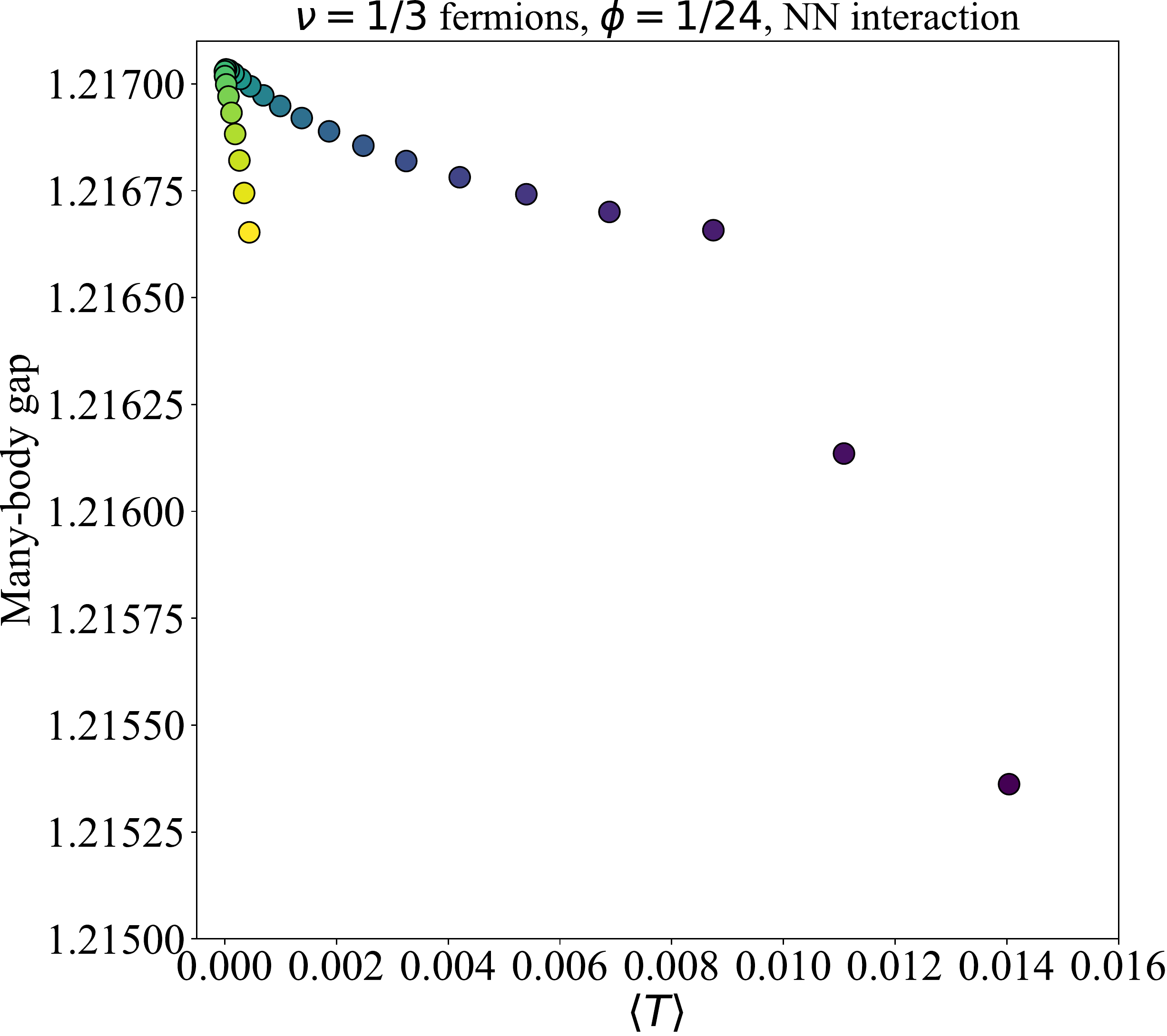}
        \includegraphics[width=2.2in]{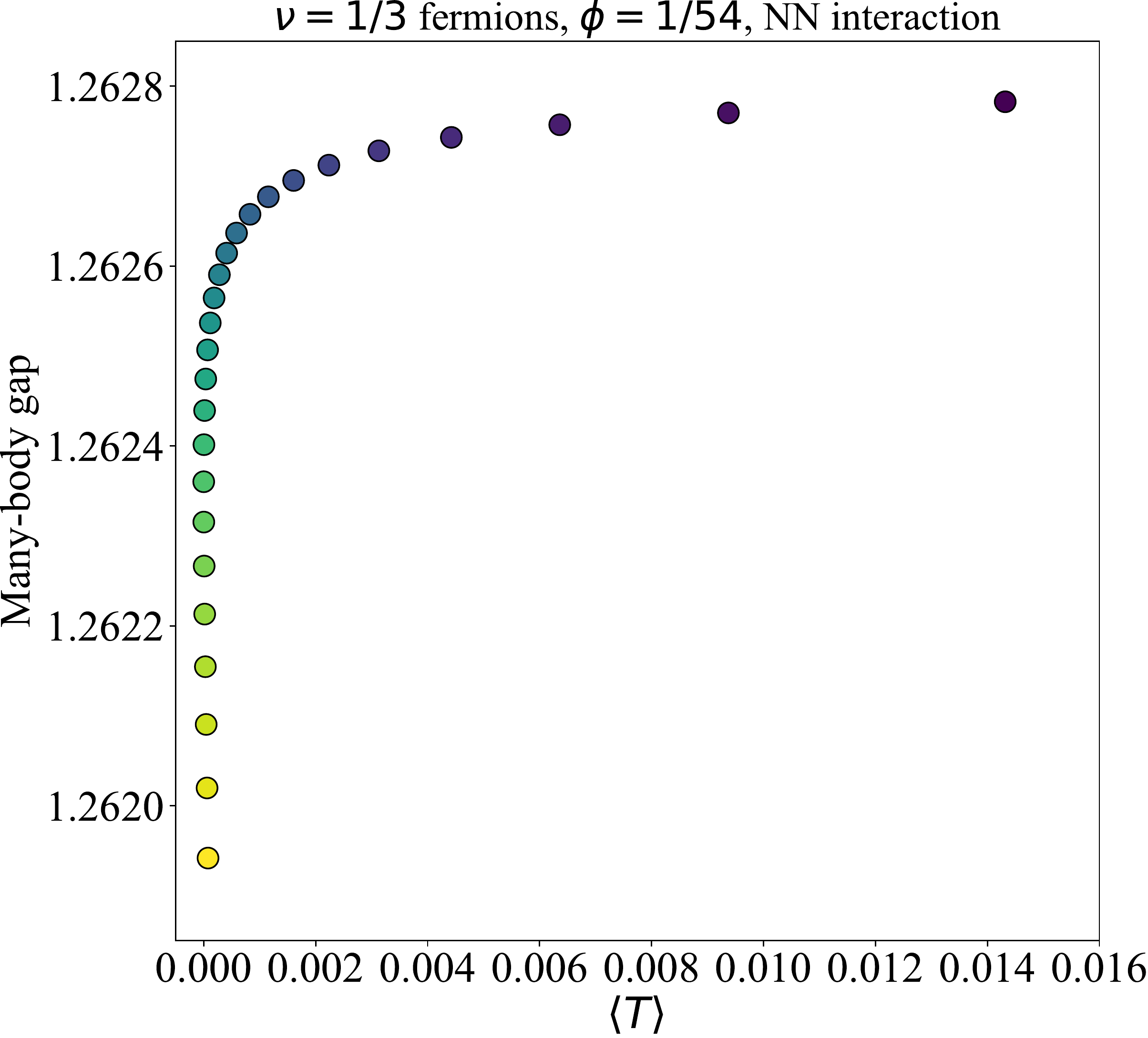}
        \includegraphics[width=2.45in]{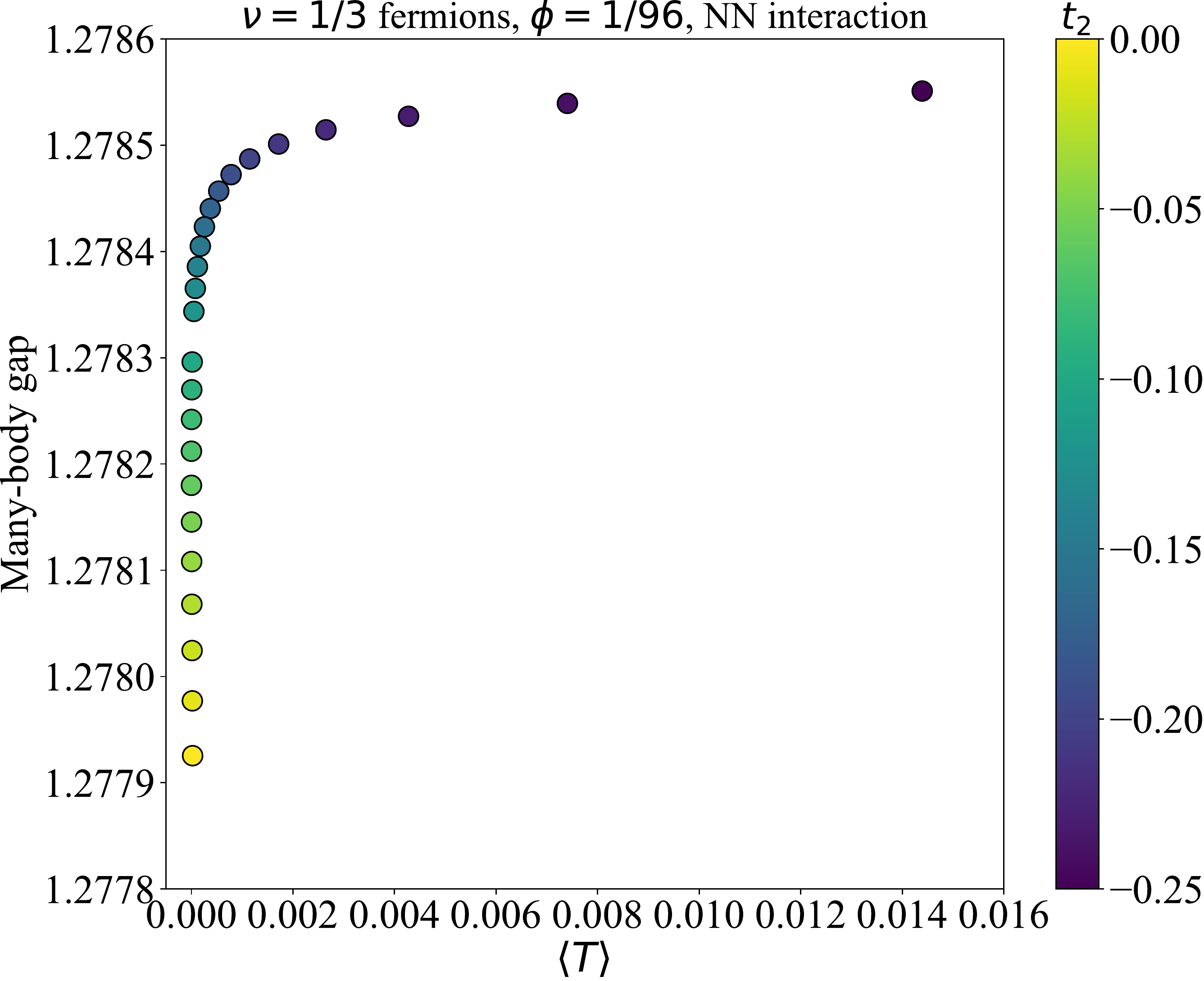}
        \caption{\label{gap-v-trace-fermions-NN} Gap between first excited state and 3-fold degenerate ground state for $N_p = 8$ fermions occupying the ground state band of the tight binding model \eqref{quartic-harper} at filling $\nu=1/3$. The interaction between fermions is nearest neighbor, density-density repulsion. We plot the dependence of the gap on the TISM $\expval{\mathcal{T}}$ as the NNN hopping parameter $t_2$ is varied. Each panel shows a different value of flux per plaquette $\phi$: from left to right we have $\phi=1/24$, $\phi=1/54$ and $\phi=1/96$.}
        \end{figure*}
        
        We have found that the form of the interaction has a significant effect on the relative stability of FQH phases in bands with different hopping parameters. Initially, we chose short-ranged interactions known to stabilize Laughlin fluids of bosons and fermions: a hard-core repulsion for bosons, and a NN repulsion for fermions (with $v=1$). We plot the gap between the first excited state and $q$-fold quasidegenerate ground states for bosons and fermions with these interactions in Figs.~\ref{gap-v-trace-bosons} and~\ref{gap-v-trace-fermions-NN}, respectively. These two cases produce quite different results in the small flux per plaquette regime. For the case of hard-core bosons, we observe the largest gap in the Hofstadter model ($t_2=0$) and the smallest gap in the zero-quadratic model ($t_2 = -1/4$), but we observe the opposite for fermions with NN repulsion.
        
        One possible explanation for this discrepancy is the commensurability between the shape of the single-particle wavefunctions and that of the interaction potential. The continuous rotational symmetry of the onsite interaction that we use for bosons may favor FQH phases in the Hofstadter model, which possess a continuous rotational symmetry in the continuum limit, whereas the $C_4$-symmetric NN interaction may favor models possessing $C_4$ symmetry but not continuous rotational symmetry in this same limit.
        
        \begin{figure*}[t]
        \includegraphics[width=2.1in]{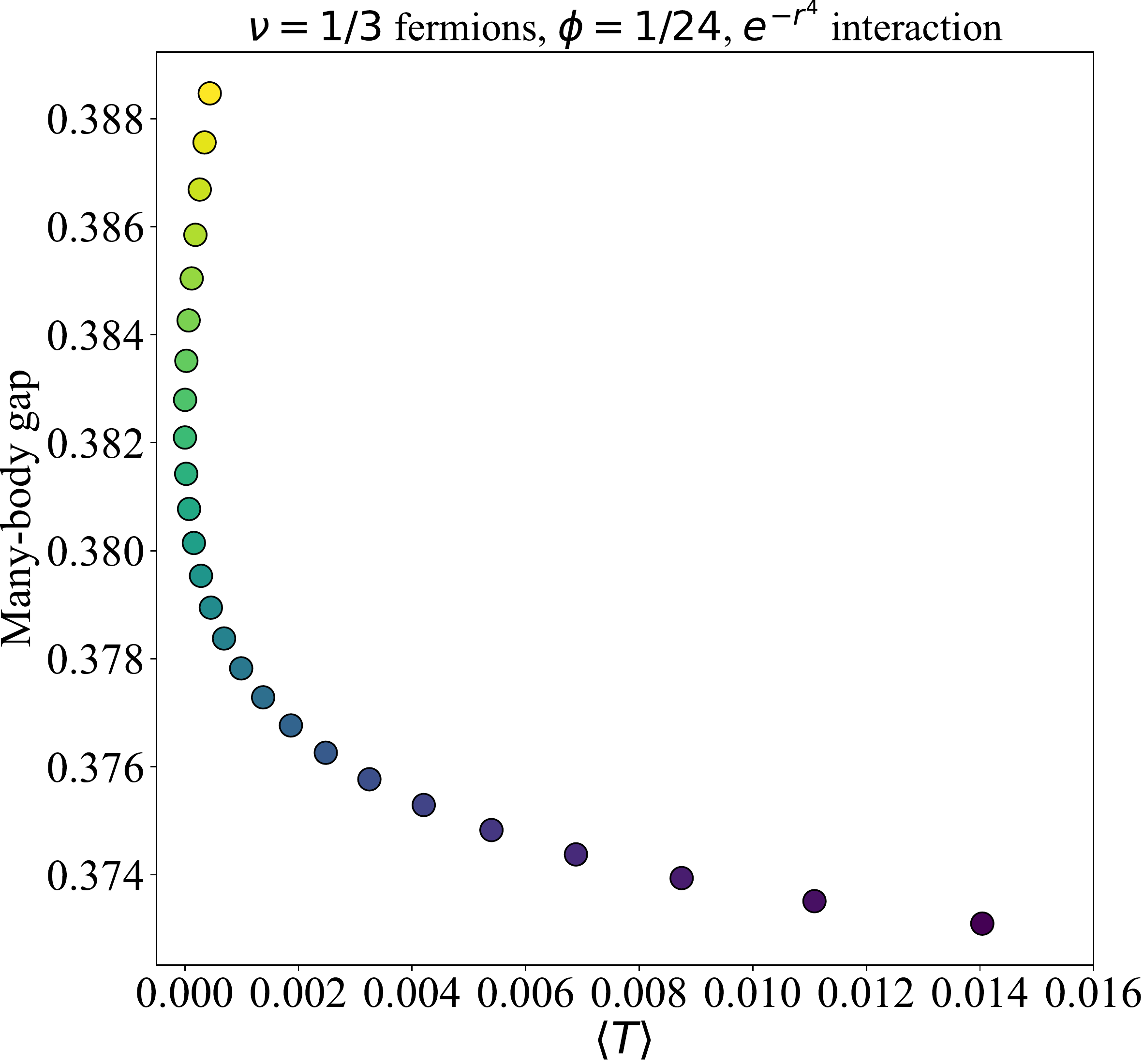}
        \includegraphics[width=2.15in]{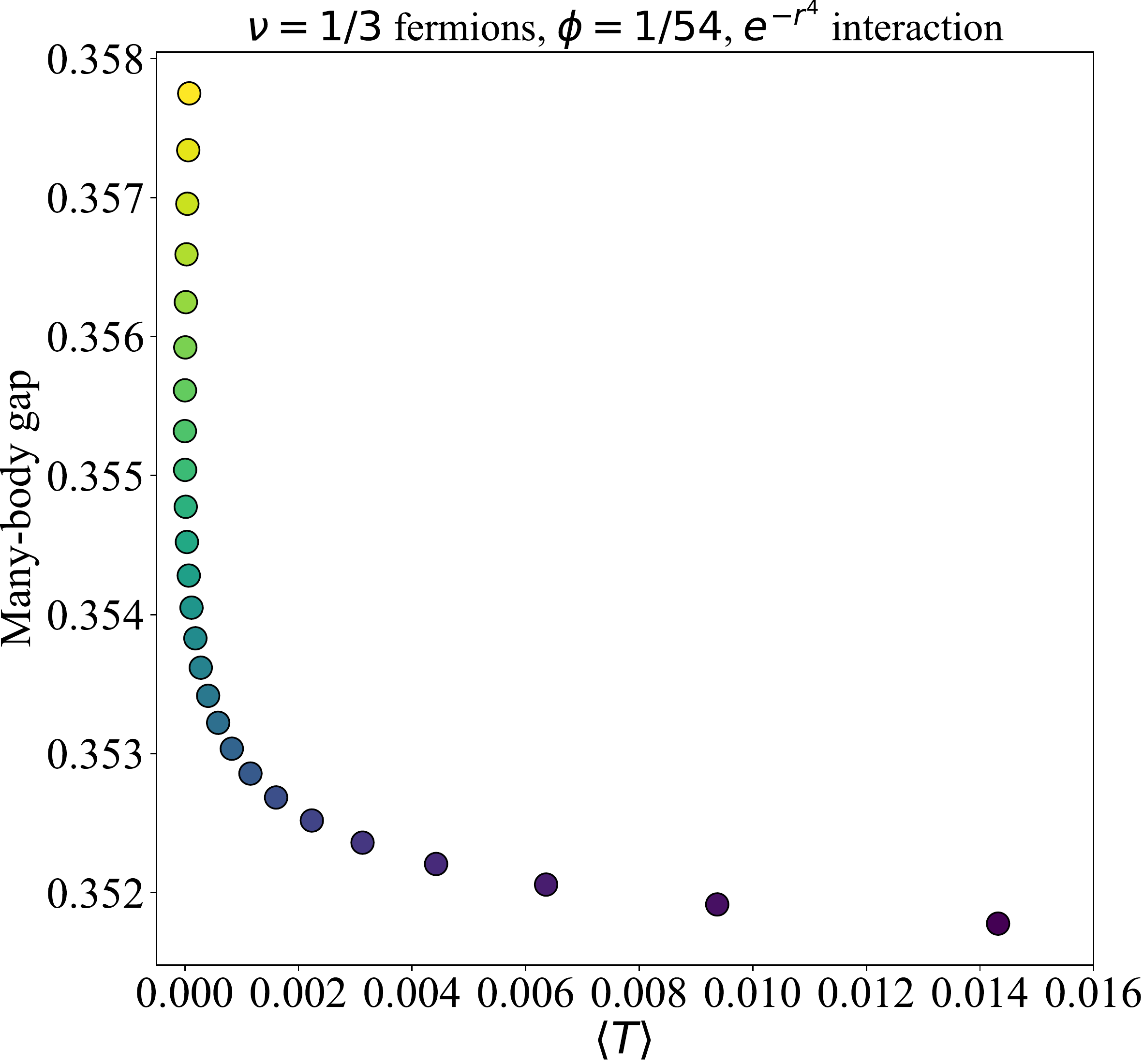}
        \includegraphics[width=2.45in]{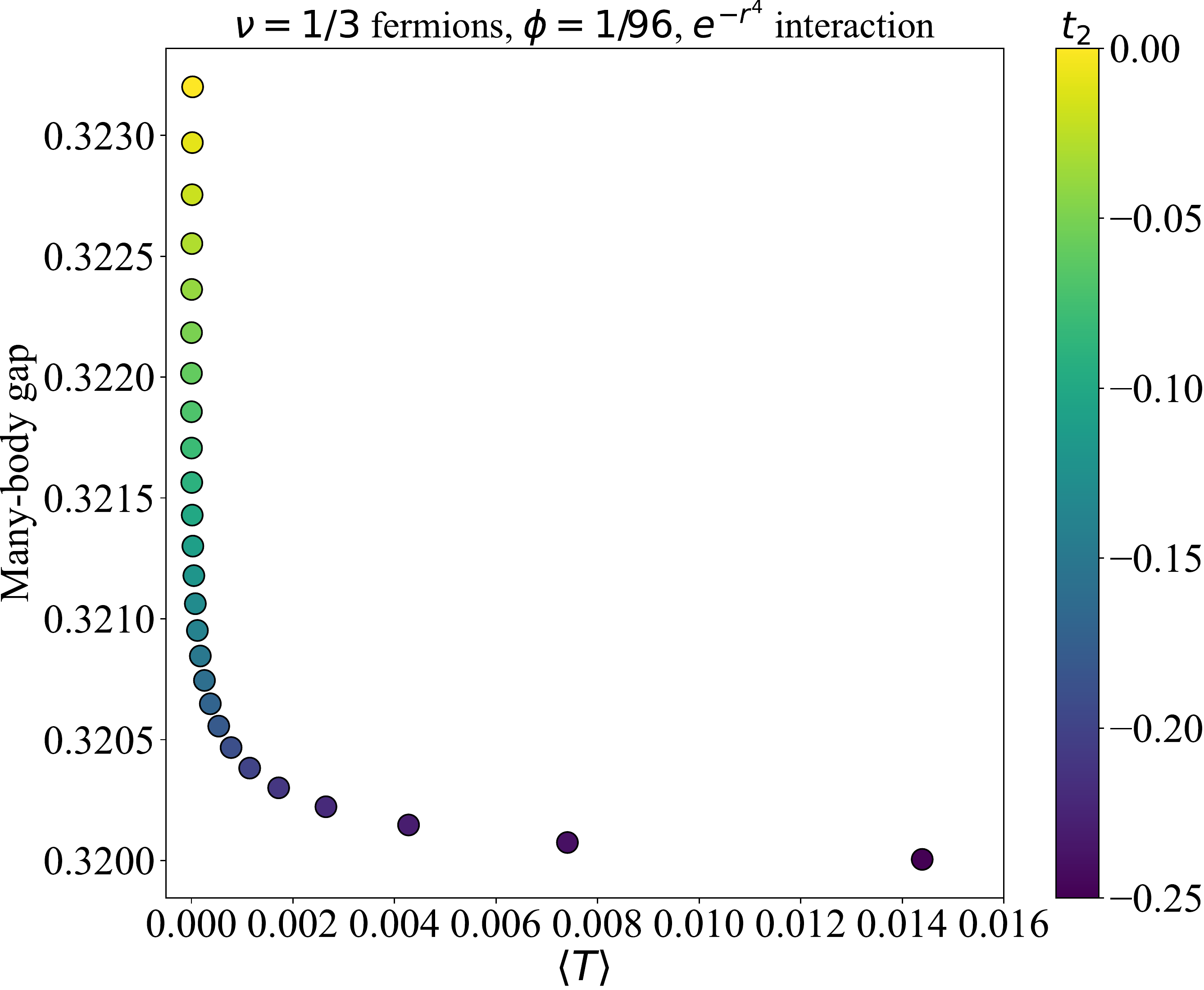}
        \caption{\label{gap-v-trace-fermions} Gap between first excited state and 3-fold degenerate ground state for $N_p = 8$ fermions occupying the ground state band of the tight binding model \eqref{quartic-harper} at filling $\nu=1/3$. The interaction between fermions is a repulsive density-density interaction that falls off as $e^{-r^4}$, as in \eqref{exp-potential}. We plot the dependence of the gap on the TISM $\expval{\mathcal{T}}$ as the NNN hopping parameter $t_2$ is varied. Each panel shows a different value of flux per plaquette $\phi$: from left to right we have $\phi=1/24$, $\phi=1/54$ and $\phi=1/96$. In contrast with the case of NN interactions, for each case the largest gap is observed for the ordinary Hofstadter model with $t_2 = 0$.}
        \end{figure*}
        
        To further explore this, we choose an interaction that depends on the squared distance between lattice sites, $|\bm{m} - \bm{n}|^2$. While such an interaction is not continuously rotationally symmetric on the lattice, it is in the continuum limit. The particular form of interaction we choose is
        \begin{align}
        \label{exp-potential}
        v_{\bm{m}\bm{n}}^{\text{exp}} = e^{-|\bm{m} - \bm{n}|^4}.
        \end{align}
        The interaction strength is short ranged, and we truncate the interaction to include up to third nearest neighbors; that is, $v_{\bm{m}\bm{n}}$ is nonzero only for $|\bm{m}-\bm{n}| \leq \sqrt{5}$. Surprisingly, modifying the interaction for fermions in this way is enough to change the dependence of the gap on $t_2$, so that $\expval{\mathcal{T}}$ now agrees well with the case of bosons with onsite interactions, as shown in Fig.~\ref{gap-v-trace-fermions}. Since this interaction is both continuous and continuously rotationally symmetric in the continuum limit, we argue that it is more relevant to experiments and favorable for continuously rotationally symmetric Landau levels in the Hofstadter model.
        
        We expect from the GSH that the TISM should provide the dominant measure of stability in the $\phi \rightarrow 0$ limit. As is apparent when comparing Fig.~\ref{trace-delta-plot} with Figs.~\ref{gap-v-trace-bosons} and~\ref{gap-v-trace-fermions}, the value of $t_2$ that minimizes $\expval{\mathcal{T}}$ does not necessarily maximize the gap. However, we do observe that, for the continuously rotationally symmetric interactions considered in Figs.~\ref{gap-v-trace-bosons} and~\ref{gap-v-trace-fermions}, the gap in the Hofstadter regime, where $\expval{\mathcal{T}}$ asymptotically vanishes as $\phi\rightarrow 0$, dominates over the zero-quadratic regime, for which $\expval{\mathcal{T}} > 0$ asymptotically, in broad agreement with the GSH.
        
        The GSH relies on the GMP algebra of projected density operators. It is therefore most relevant to problems where the interactions are of a density-density type and purely functions of the separation, i.e., isotropic in the continuum. On the lattice, rotational symmetry is necessarily broken, but the extent to which the lattice interactions mimic a continuum isotropic interaction should govern the applicability of the GSH. In light of these observations, it is unsurprising that the TISM is correlated with the size of the many-body gap in the models with an isotropic continuum analogue.

    \section{Conclusion}
    \label{sec-conclusion}
    
        In conclusion, we have constructed tight-binding models based on the Harper-Hofstadter model that have non-Landau level flat bands as their eigenstates in the continuum limit. We have shown that the geometry of these bands is measurably distinct from that of Landau levels, even when the dispersion, Berry curvature, and FS metric can be treated as flat. We also demonstrated how this is reflected in their finite-wavevector electromagnetic responses to an applied electric field. We introduced a transfer matrix and studied the electronic delocalization transition in the lowest level of the zero-quadratic model with quenched onsite disorder and found a localization-length critical exponent of $2.57(3)$, which is in close agreement that of the disordered Hofstadter model $2.58(3)$~\cite{puschmann2019integer}, and in concurrence with localization-length exponents for Landau levels and the IQHE~\cite{slevin2009critical,obuse2012finite,gruzberg2017geometrically,klumper2019random,lutken2007geometric,lutken2019elliptic,puschmann2019integer,ippoliti2018integer,zhu2019localization,sbierski2020criticality,huang2020numerical,puschmann2021green}. This provides evidence that supports a broad universality of the IQHE localization-delocalization critical exponent independent of specific model parameters. Finally, we diagonalized interactions projected to these bands, found signatures of $\nu=1/2$ and $1/3$ Laughlin states, and highlighted the role of the FS metric and form of interparticle interactions on FCI stability.
        
        The results in this article find applications in various aspects of current research. For example, two-dimensional moir\'e lattices, such as magic-angle twisted bilayer graphene (MATBG), have recently emerged as a new platform for realizing and studying strongly interacting systems with topological flat bands~\cite{suarez_morell_flat_2010,bistritzer_moire_2011,moon_energy_2012, andrews2020tbg}. MATBG in particular has been observed to exhibit superconductivity~\cite{cao_unconventional_2018,yankowitz_tuning_2019,lu_superconductors_2019}, charge ordering~\cite{jiang_charge_2019}, and ferromagnetism leading to the formation of Chern states~\cite{sharpe_emergent_2019,chen_tunable_2020,wu_chern_2021}. FCIs have also been predicted to occur~\cite{abouelkomsan_particle-hole_2020,ledwith_fractional_2020,repellin_chern_2020}, and have recently been experimentally observed via compressibility measurements in the presence of a perpendicular magnetic field~\cite{Spantoneaan8458,xie_fractional_2021}. Other experimental works have identified single-particle Chern bands in the FCI regime~\cite{Jotzu2014,Aidelsburger:2014hm,Aidelsburger:2013ew}, which can also be realized in periodically-driven quantum or Floquet systems with a Berry curvature that can be engineered and measured~\cite{Flaschner1091}. These experimental studies of FCIs necessitate a theoretical understanding of bands that are not Landau levels yet host fractionalized topological states, as well as general criteria for FCI stability.
        
        Our results demonstrate the importance of the saturation of the determinant and trace inequalities in their domain of applicability. Yet, at the same time, they also highlight the limitations of the GSH. In particular, the geometry of the interactions are seen to play a crucial role in whether the measures inferred from the GSH predict the size of the gap and thus the stability of the FCI phases. Extensions of the GSH, such as the recent proposal by Simon and Rudner~\cite{simon2020gsh}, may address the sorts of interactions that we have studied here and could be an interesting direction for future work. Other avenues for future research include investigating alternative models that host stable FCIs despite being dissimilar to Landau levels in the continuum, as well as their quantum phase transitions~\cite{Jian20} and effective field theory descriptions~\cite{Sohal18}. Moreover, insight into the stability of FCIs continuously connected to higher non-Landau levels or cases where there is significant non-Landau level mixing is still an open problem. We hope that the family of models introduced in this paper will inspire further study in these areas.
    
    \begin{acknowledgments}
        The authors thank Tom Jackson for his band geometry code and for collaboration on related work. We also thank authors of the \textsc{DiagHam} package, which was used in this work. D.B., F.H., and R.R. acknowledge support from the NSF under CAREER Grant No. DMR-1455368; S.T. acknowledges support from the NSF under GRFP Grant No. DGE-1845298; B.A. and R.R. acknowledge support from the University of California Laboratory Fees Research Program funded by the UC Office of the President (UCOP), grant ID LFR-20-653926; and all authors thank the Mani L. Bhaumik Institute for Theoretical Physics.
    \end{acknowledgments}

    \appendix
    
    \section{Weak-field effective Hamiltonian for the Hofstadter model with \texorpdfstring{$C_4$}{C4}-symmetric NNN hopping}
    \label{allNNN}
    
        For completeness, we reproduce here the Hofstadter model with all NNN hopping terms that respect $C_4$ symmetry, including the hopping diagonally across a plaquette. The tight-binding Hamiltonian is
        \begin{align}
        H_0 = &-t_1 \left(T_x + T_y\right)\nonumber - t_2 \left(T_x^{2} + T_y^{2}\right)\\ &- t_3 \left(T_xT_y + T_y T_x\right) + \text{H.c.}.
        \end{align}
        
        We can write this in terms of the generators $K_a$ as
        \begin{align}
        H_0 = &-2t_1 \left[\cos\left(K_x\right) + \cos\left(K_y\right)\right]\nonumber\\ &-2t_2 \left[\cos\left(2K_x\right)+\cos\left(2K_y\right)\right]\nonumber\\ &- 4t_3 \cosh\left(\frac{\phi}{2}\right)\cos\left(K_x + K_y\right).
        \end{align}
        The factor of $\cosh(\phi/2)$ is non-universal and results from our ordering prescription. Expanding to quartic order, we obtain the small-$\phi$ effective Hamiltonian
        \begin{align}
        H_{\text{eff}} = h_{ab}K_a K_b + \lambda_{abcd} K_a K_b K_c K_d 
        \end{align}
        with coefficients
        \begin{align}
        h_{11} = h_{22} &= t_1 + 4t_2 + 2t_3,\\
        h_{12} &= 2t_3,
        \end{align}
        and 
        \begin{align}
        \lambda_{1111} = \lambda_{2222} &= -\frac{1}{3}\left(\frac{t_1}{4} + 4t_2 + \frac{t_3}{2} \right),\\
        \lambda_{1112}  = \lambda_{1222} &= -t_3/3,\\
        \lambda_{1122} &= -t_3/2.
        \end{align}
    
    \section{Zero-energy states for the \texorpdfstring{$a^4 + a^{\dag 4}$}{a4+ad4} Hamiltonian}
    \label{zero-enegy-quartic}
        
        In this section, we obtain an analytic expression for the lowest energy wavefunction of a specific quartic model. 
        
\begin{figure}
        \includegraphics[width=\linewidth]{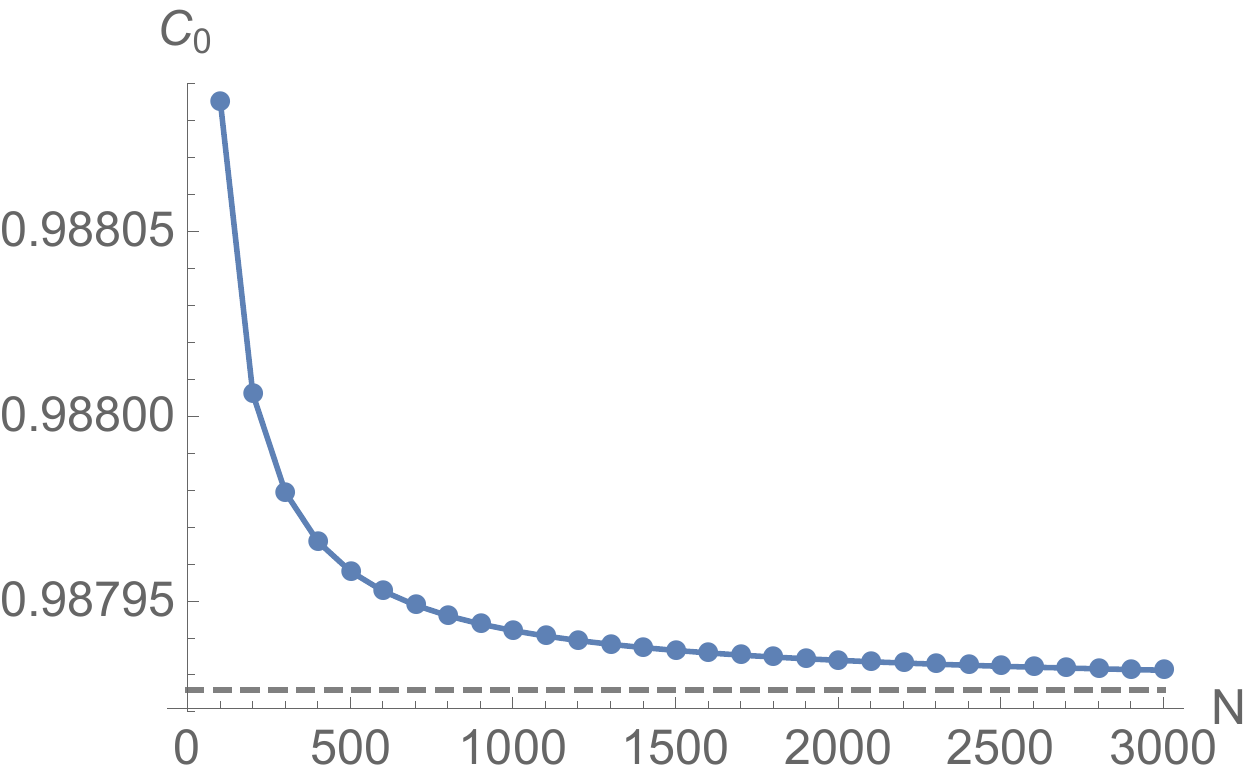}
        \caption{\label{fig:C0} Convergence of $C_0$ with the number of iterations $N$. The exact value $C_0=0.987926$ is marked with a dashed line.}
        \end{figure}

        We take the Hamiltonian
        \begin{align}
        \hat{H}=\omega\left[a^4 +\left(a^\dagger\right)^4\right]
        \end{align}
        and write a general wavefunction
        \begin{align}
        \ket{\psi}=\sum_n C_n\ket{n}
        \end{align}
        as a sum over Landau level states. The eigenvalue equation becomes
        \begin{align}
        \begin{split}
        \hat{H}\ket{\psi}=\omega\sum_n&C_n[\sqrt{n(n-1)(n-2)(n-3)}\ket{n-4}\\
        &+\sqrt{(n+1)(n+2)(n+3)(n+4)}\ket{n+4}]\\
        =E\sum_n &C_n\ket{n}.
        \end{split}
        \end{align}
        Taking inner products with different Landau level states yields
        \begin{align}
        \omega\left[C_{4(p+1)}\alpha_{-4}(4(p+1))+C_{4(p-1)}\alpha_{+4}(4(p-1))\right]=EC_{4p},
        \end{align}
        where we have defined 
        \begin{align}
        \alpha_-(n)&=\sqrt{n(n-1)(n-2)(n-3)}\\
        \alpha_+(n)&=\sqrt{(n+1)(n+2)(n+3)(n+4)}.
        \end{align}
        Now we seek to obtain a state with energy $E=0$. Substituting this into some of the coefficient equations, we find
        \begin{align}
        C_4 &=0,\\
        C_8&=-C_0\frac{\alpha_{+4}(0)}{\alpha_{-4}(8)},\\
        C_{12} &=0,\\
        C_{16} &=-C_{8}\frac{\alpha_{+4}(8)}{\alpha_{-4}(16)}\\
        &=C_{0}\frac{\alpha_{+4}(8)\alpha_{+4}(0)}{\alpha_{-4}(16)\alpha_{-4}(8)}.
        \end{align}
        The pattern of coefficients continues in the same manner. We define
        \begin{align}
        \beta(n)=\frac{\alpha_{+4}(8(n-1))}{\alpha_{-4}(8n)}
        \end{align}
        so that
        \begin{align}
        C_{8}&=-\beta(1)C_0\\
        C_{16}&=+\beta(2)\beta(1)C_0\\
        &\vdots\nonumber\\
        C_{8p}&=(-1)^pC_0\prod_{k=1}^{p}\beta(k)\\
        &\equiv\gamma(p)C_0,
        \end{align}
        where we have defined $\gamma(p)$ in the final line. Mathematica simplifies the function $\gamma$ to
        \begin{align}
        \gamma(p)=&\frac{\sqrt[4]{\frac{2}{\pi}} \Gamma \left(\frac{7}{8}\right)}{\Gamma \left(\frac{1}{8}\right)} \times\nonumber \\
        &\sqrt{\frac{\Gamma \left(p+\frac{9}{8}\right) \Gamma \left(p+\frac{5}{4}\right) \Gamma \left(p+\frac{11}{8}\right) \Gamma \left(p+\frac{3}{2}\right)}{ \Gamma \left(p+\frac{13}{8}\right) \Gamma \left(p+\frac{7}{4}\right) \Gamma \left(p+\frac{15}{8}\right) \Gamma (p+2)}},
        \end{align}
        with 
        \begin{align}
        \Gamma(z)=\int_0^\infty x^{z-1}e^{-x}\dd x.
        \end{align}
        We now have both a recursive and a non-recursive relation for the coefficient $C_{8p}$ (with all other $C_k=0$). The complete wavefunction is
        \begin{align}
        \ket{\psi}&=\sum_nC_n\ket{n}\nonumber\\
        &=C_0\sum_n\gamma(n)\ket{n},
        \end{align}
        which leads to the normalization condition
        \begin{align}
        1=\left|C_0\right|^2\left[1+\left|\gamma(1)\right|^2+\left|\gamma(2)\right|^2+\ldots\right].
        \end{align}
        Using Mathematica, we find
        \begin{align}
        \frac{1}{\left|C_0\right|^2}=_4F_3\left(\left[\frac{1}{8},\frac{1}{4},\frac{3}{8},\frac{1}{2}\right],\left[\frac{5}{8},\frac{3}{4},\frac{7}{8}\right],1\right),
        \end{align}
        where $_4F_3(a;b;z)$ is a generalised hypergeometric function. Solving for $C_0$, yields $C_0=0.987926$, which agrees well with numerics, as shown in Fig.~\ref{fig:C0}.
        
    \section{Variation of the localization-delocalization exponent with NNN hopping strength}
    \label{exponentNNN}
    	
    	In the body of the paper, we found that the electronic localization-delocalization exponent in the zero-quadratic model with quenched onsite disorder is equal to that of the disordered Hofstadter model. One may well ask whether this exponent is universal to all values of NNN hopping strength $t_2$ since $t_2=-1/4$ is the fine-tuned value where quadratic terms vanish. In this appendix we present the results of varying $t_2$ on the electronic localization-delocalization exponent.
    	
    	For values of $t_2$ around $t_2=-1/4$, at $\phi=(1/10)(\phi_0/2\pi)$ for strip widths of  $12,16,24,32,48,64,96,$ and $128$ sites with periodic boundary conditions, 51 energies were selected linearly spaced in the neighborhoods of the critical energies of the lowest-levels.
    	For the disorder potential, we used uncorrelated white noise taken from a uniform distribution, $V_{\ell,w}=\mathrm{uniform}(-s/2,s/2)$ with strength $s=0.1t_1$, a value which is large compared to the intrinsic level widths of the levels respectively, but smaller than the level spacings.
    	We did this for 2 realizations of strips of length $L=10^5$, for a cumulative strip length of $L=2\times10^5$.
    	
    	\begin{figure}[thp]
    		\includegraphics[width=1\linewidth]{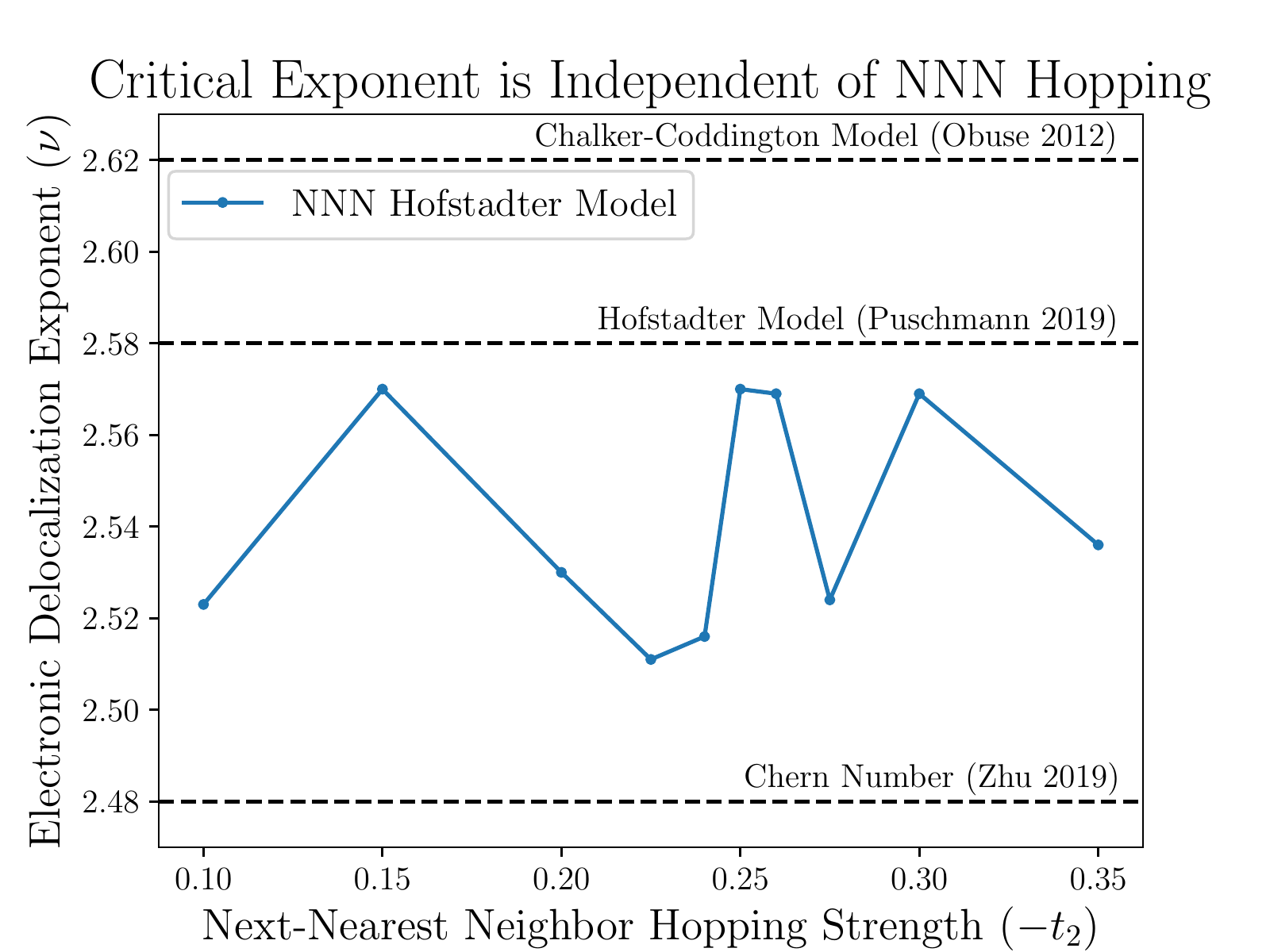}
    		\caption{\label{fig:variation} Electronic localization-delocalization critical exponent $\nu$ as dependent on NNN hopping strength $t_2$. We see that $\nu$ is independent of $t_2$ up to the precision of the simulations as all exponents are in the interval $(2.51,2.57)$, and we estimate a precision of $\pm 0.03$ on each exponent determined. The determined critical exponents are in concurrence with recent theoretical predictions of the IQHE localization-delocalization critical exponent.}
    	\end{figure}
    	
    	We find that the electronic localization-delocalization critical exponent is independent of the NNN hopping strength, and that the critical exponent is consistent with recent theoretical predictions of the IQHE localization-delocalization critical exponent~\cite{obuse2012finite,puschmann2019integer,zhu2019localization}, as shown in Fig.~\ref{fig:variation}. In particular, the mean value of the exponents as dependent on $t_2$ is $2.54(3)$, with all exponents in the interval $(2.51,2.57)$. We estimate that the numerically estimated exponents are accurate to within a precision of $\pm 0.03$.
    	
    	This mean value is slightly smaller than that of the disordered Hofstadter model as determined by Puschmann~et al.~with widths of up to $W=768$ sites~\cite{puschmann2019integer}.
    	We note that the largest widths studied here are $W=128$ sites so it is possible that finite-size effects lead to the difference in exponents. In the disordered Hofstadter model it appears that finite-size effects result in a smaller value of the critical exponent. Therefore we believe that the value of the critical exponent for the disordered zero-quadratic model may only be smaller than the value of the disordered Hofstadter model as a result of finite-size effects.
    
    \bibliographystyle{apsrev4-2}
%

\end{document}